\documentclass[reprint,showpacs,superscriptaddress,amsmath,amssymb,aps]{revtex4-1}

\usepackage{graphicx}
\usepackage{color}
\usepackage{dcolumn}
\usepackage{bm}
\usepackage{amsmath}
\usepackage{amsfonts}
\usepackage{amssymb}
\def\D{\mathrm{d}}

\begin{document}

\title{Capacitance and compressibility of heterostructures \\
with strong electronic correlations}

\author{Kevin Steffen}
\altaffiliation{Kevin.Steffen@physik.uni-augsburg.de} 
\affiliation{Center for Electronic Correlations and Magnetism, EP VI, Institute of Physics, University of Augsburg, 86135 Augsburg, Germany}
\author{Raymond Fr\'esard}
\affiliation{Normandie Universit\'e, ENSICAEN, UNICAEN, CNRS, CRISMAT, 14050 Caen, France}
\author{Thilo Kopp}
\affiliation{Center for Electronic Correlations and Magnetism, EP VI, Institute of Physics, University of Augsburg, 86135 Augsburg, Germany}

\begin{abstract}
Strong electronic correlations related to a repulsive local interaction suppress the electronic compressibility in a single-band model, and the capacitance of a corresponding metallic film is directly related to its electronic compressibility. Both statements may be altered significantly when two extensions to the system are implemented which we investigate here: 
(i) we introduce an attractive nearest-neighbor interaction $V$ as antagonist to the repulsive on-site repulsion $U$, and (ii) we consider nano-structured multilayers (heterostructures) assembled from two-dimensional layers of these systems. We determine the respective total compressibility $\kappa$ and capacitance $C$ of the heterostructures within a strong coupling evaluation, which builds on a Kotliar-Ruckenstein slave-boson technique. Whereas the capacitance $C(n)$ for electronic densities $n$ close to half-filling is suppressed---illustrated by a correlation induced dip in $C(n)$---it may be appreciably enhanced close to a van Hove singularity. 
Moreover, we show that the capacitance may be a non-monotonic function of $U$ close to half-filling for both attractive and repulsive $V$.
The compressibility $\kappa$ can differ from  $C$ substantially, 
as $\kappa$ is very sensitive to internal electrostatic energies which in turn depend on the specific set-up of the heterostructure.
In particular, we show that a capacitor with a polar dielectric has a smaller electronic compressibility and is more stable against phase separation than a standard non-polar capacitor with the same capacitance.
\end{abstract}

\date{\today}

\maketitle

\section{\label{sec:introduction}Introduction}
The experimental and theoretical investigation of oxide heterostructures has advanced significantly in recent years, not least with the discovery of a metallic state at the interface of two bulk band insulators, LaAlO$_3$ and SrTiO$_3$~\cite{Ohtomo2004}. It is the electronic reconstruction in the vicinity of the interface, which 
generates the metallic state~\cite{Nakagawa2006,Thiel2006} and allows for the coexistence of superconductivity and magnetism~\cite{Li2011a,Moler2011}.
Furthermore, it is suggested to cause the formation of 
phase-separated states~\cite{Ariando2011,Grilli2012,Pavlenko2013,Scopigno2016}; even a state with negative electronic compressibility at the interface has been identified~\cite{Li2011b,Tinkl2012}.

Electronic devices are generically heterostructures where functionalities are determined by interfaces and surfaces. The conventional semiconductor physics was fascinatingly successful in this respect. Heterostructures with transition metal oxides may come to rival the silicon based systems provided that electronic mobilities can be sustained sufficiently high~\cite{Mannhart2010a}. 
The recent progress in the fabrication of multilayered systems makes it possible to manufacture stable polar nano-structures with physical properties that allow to functionalize oxide interfaces and surfaces~\cite{Noguera2008}.

One of the advantages of oxide heterostructures is the possibility to generate strongly correlated electronic states~\cite{Freericks2016}. High-temperature superconductivity, the transition to a Mott insulating state as well as exotic magnetism are particular manifestations of strong electronic correlations. What is  the impact of strong correlations on the electronic reconstruction in heterostructures? In this paper we focus on the electronic compressibility controlled by the electronic density at the respective interfaces and surfaces. The density at interfaces can be tuned by gate biases, in fact by at least an order of magnitude in the case  of 
LaAlO$_3$/SrTiO$_3$. 
Moreover, as the electronic compressibility is intimately related to the capacitance, the question arises how the capacitance is controlled by strong electronic interactions---in dependence on the electronic density.

Long-range Coulomb interactions generate a negative compressibility in a dilute homogeneous electron system. Corresponding analytical calculations in perturbation theory are supported by the results of quantum Monte Carlo evaluations~\cite{Ceperley1978,Tanatar1989,Bulutay2002}. It is mostly the exchange interaction which is responsible for this behavior. Electronic exchange produces a negative compressibility on account of the formation of exchange holes, and capacitances with electrodes comprising such a dilute electron systems may be enhanced well beyond their geometrical capacitance value~\cite{Eisenstein1992,Eisenstein1994,Kopp2009,Li2011b}. For the case of very dilute electron systems, Wigner crystallization is suggested to explain observed capacitance enhancements~\cite{Shklovskii2010}.

In this work we want to investigate systems for intermediate filling, that is, we do not consider the dilute homogeneous electron system where exchange becomes so pronounced. We focus on a single-band lattice system and consider only on-site and nearest-neighbor interaction, and evaluate compressibility and capacitance of the electronic system. The capacitance of interacting electrons in a ring geometry pierced by a magnetic flux was introduced by B{\"u}ttiker~\cite{Buettiker1994}, and the flux-dependent capacitance was evaluated for a disordered electronic system with electron interactions represented by an extended Hubbard model~\cite{Mukherjee2011}. Here we will not consider inhomogeneous systems; this is beyond the scope of this work. Inhomogeneities are introduced exclusively through the multilayer setup of the planar heterostructure.

An investigation of the capacitance of multilayers with strongly correlated materials has been conducted recently by Hale and Freericks~\cite{Freericks2012}. They considered a heterostructure with non-interacting leads however with a barrier represented by a Falicov-Kimball model~\cite{Falicov1969}, and they identify an enhancement in the capacitance with increasing thickness of the barrier~\cite{Freericks2012,Freericks2016}. We evaluate a very different, rather complementary set-up: we build the capacitance from a pair of two-dimensional (2D) metallic electrodes which 
for themselves represent correlated electron systems and the barrier in between is a polar or non-polar dielectric material. 

In these systems with sizable band filling, the long-range Coulomb interaction is screened within the metallic plates and the exchange does not play a dominant role in the compressibility. Instead, short-range interactions control the compressibility---apart from the kinetic term. Also further one-particle contributions such as spin-orbit coupling~\cite{Grilli2012,Bucheli2014,Ste15,Seibold2015} or transverse potential profiles~\cite{Scopigno2016} can have an impact on the compressibility. We will briefly address systems with Rashba spin-orbit coupling in the last part of Sec.~\ref{sec:layout}.

The repulsive on-site Coulomb interaction is screened by the polarization of neighboring atoms, for example in the cuprates by the polarization of oxygen ions~\cite{Boer84,Brink97}. The screening depends on the excitations of the local cluster of atoms. It can be quite different for the nearest-neighbor interaction term and produce an attractive nearest-neighbor interaction, as in certain iron-pnictides~\cite{Saw09,Ber09}. Polaronic effects from strong electron-lattice coupling may enhance an attractive nearest-neighbor interaction~\cite{Mic90}. For LaAlO$_3$/SrTiO$_3$ polaronic effects have been observed~\cite{Strocov2016} and investigated within a first-principles-based scheme~\cite{Junquera2016}.

A repulsive on-site interaction $U$ will be effective close to half-filling and it will generically lead to a filling-dependent suppression of the compressibility. As antagonist to the repulsive on-site interaction we introduce an attractive nearest-neighbor interaction $V<0$.  The nearest-neighbor interaction $V$ can generate a charge instability~\cite{Fre16}, even for large $U$ if the electronic density is not too close to half-filling. 
A nearest-neighbor repulsive interaction, on the other hand, is expected to rather enhance the suppression of the compressibility through the on-site Coulomb repulsion. It is only for the heterostructures with dilute electronic systems~\cite{Kopp2009} that the exchange term from the repulsive non-local interaction increases the compressibility significantly. The sign of the nearest-neighbor interaction has to be determined from a careful microscopic
analysis of the effective (screened or even overscreened) interactions in the heterostructure. Such an analysis is not the scope of the present work; we focus on the similarly intriguing question how electronic correlations modify the capacitance of the heterostructure for specified interaction terms with scales
$U$ and $V$.

Whereas the (inverse) electronic compressibility is related to the second derivative of the free energy with respect to the total number of mobile electrons, the capacitance is related to the difference in electronic charge of subsystems. These subsystems are characterized by an electrochemical potential difference (voltage), which is the conjugate thermodynamic variable with respect to the difference in electronic charges. The two 
density-response functions (compressibility and differential capacitance) appear to be very similar and, in fact, the standard experiment to measure the electronic compressibility is the determination of capacitances.

The positivity of the compressibility is a stability criterion---as well as the positivity of the capacitance of the heterostructure.  As the chemical potential is the first derivative of the free energy with respect to the total number of mobile electrons, a negative electronic compressibility would signify that an increase in the electronic charge entails a decrease in the chemical potential. The electronic compressibility may attain a negative value even in an equilibrium state as long as this is compensated by a sufficiently large positive value from the ionic system. Moreover, it is to be emphasized that electronic subsystems may well display negative compressibility but the total electronic system still has a positive compressibility on account of a large electrostatic contribution, even for nanosized lateral extension of a capacitive electronic system. Here, we will discuss that in systems with polar dielectric materials this latter situation is often realized due to the electrostatic energy of the polar layers. Therefore, a scenario where a planar substructure of a heterostructure has negative electronic compressibility does not necessarily imply that the system is instable and develops a phase separation---the negative compressibility rather supports a capacitance enhancement. An instructive example is the heterostructure with Rashba spin-orbit coupling in a metallic interface plane. 

Eventually, an important issue arises in this respect: when is the electronic compressibility equivalent to the capacitance of a heterostructure and when does it differ appreciably? We will deal with this question by ``deforming'' heterostructures continuously and thereby approach different geometric limits such as a standard two-plate capacitance or a set-up known from a gated LaAlO$_3$/SrTiO$_3$ multilayer structure. Thereby we will identify a charge-transfer function which represents the increase in charge of a subsystem with the increase of the total electronic charge. This charge-transfer function parameterizes the discrepancy between differential capacitance and compressibility and allows to extract information about the stability of the electronic system. We will find that heterostructures with polar dielectrics acquire the highest electronic stability, in comparison to different  non-polar heterostructures with identical capacitances.

The paper is organized as follows.  In Sec.~\ref{sec:technique} we present the extended Hubbard model and the 
Kotliar-Ruckenstein slave-boson technique to investigate correlated electron systems. In particular we compare the free energy and compressibility of the slave-boson evaluation with Hartree-Fock results in the regime of weak correlations. In Sec.~\ref{sec:basics} the basic setup for a capacitance of a polar heterostructure is introduced for which the compressibility and capacitance are determined in the strong coupling regime. 
The dependence of the capacitance on the on-site interaction $U$ is
comprehensively discussed in Sec.~\ref{sec:largeU}. Eventually, we present distinct layouts of capacitances in polar structures in Sec.~\ref{sec:layout}
and compare capacitor setups which have equal capacitances but unequal compressibilities. The results are discussed for strongly correlated systems and, finally, for electron systems subject to Rashba spin-orbit coupling. In the conclusions we readdress the distinction between the
capacitance and the compressibility of a multilayered electronic system and summarize our findings.

Weak coupling results for the compressibility
and capacitance are shortly discussed in Appendix~\ref{appendix_A}; they are markedly at variance with the strong coupling results. In 
Appendix~\ref{appendix_B} we briefly address heterostructures where the surface electrode comprises an uncorrelated electron system and the interface electrode a strongly coupled system as these configurations may often be realized in experimental setups. 
Appendix~\ref{appendix_C} presents the interrelation of the total compressibility of a structure and its respective capacitance that was used
in the sections of the main text.


\section{\label{sec:technique}Technique to investigate correlated electron systems}

In our analysis of the compressibility of a heterostructure, which is composed of two 2D metallic subsystems and further insulating layers, we resort to a technique that we already tested on
a 2D extended Hubbard model in a previous work~\cite{Fre16}, the Kotliar and Ruckenstein slave boson technique.

\subsection{\label{sec:slaveboson}Kotliar-Ruckenstein slave bosons}

Extending Barnes' pioneering work~\cite{Bar76}, slave boson representations to the most prominent correlated electron models have been introduced and studied (for a review see Ref.~\onlinecite{FKW}). Most relevant to the Hubbard model are the Kotliar and Ruckenstein (KR) representation~\cite{Kot86}, as well as its rotationally invariant generalizations~\cite{Li89,FW}. All of them possess an internal gauge symmetry group. It may be made use of to simplify the problem and to gauge away the phases of some
bosons by promoting all constraint parameters to fields~\cite{FW}. Thereby only slave boson fields remain~\cite{Fre01}, that may not Bose condense. In fact, their exact expectation values are generically non-vanishing~\cite{RFTK12} (see Ref.~\cite{Kop07} in the case of Barnes' representation to the single impurity Anderson Model), and may be approximately obtained through the saddle-point approximation that we are going to use below. The latter has been shown to compare favorably with numerical simulations in a number of cases. For instance, the comparison of the ground state energy to Quantum Monte Carlo (QMC) data on the square lattice for $U=4t$ yields an agreement in the less than 3\% range~\cite{Fre91}. On the honeycomb lattice, the location of the metal-to-insulator transition was found to be in excellent agreement with QMC data~\cite{Doll3} (for more examples see Ref.~\onlinecite{Lhou15}).

Regarding fluctuations at one loop order, RPA forms have been obtained in the low frequency-long wavelength limit, that even agree quantitatively with perturbation theory to lowest order in $U$, in both spin and charge channels~\cite{Li91}. In the complementary equal-time limit, quantitative agreement to QMC charge structure factors has been demonstrated for intermediate coupling~\cite{Zim97}.

It has recently been shown that the spin rotation invariant representation can be extended to incorporate longer range interactions~\cite{Lhou15}. 
In particular, the one-loop contribution to the compressibility has been shown to be independent of the precise structure of the screened Coulomb interaction $V_{k}$, as the latter only enters through its $k=0$ Fourier component $V_{k=0}$. This applies to the saddle-point equations as well. While it allowed to reveal several instabilities, both on the cubic~\cite{Lhou15} and square~\cite{Fre16} lattices, it has not been put to such numerical tests. In fact, it is only established to be variationally controlled in the large dimensionality limit~\cite{Lhou15}. Below, we show that, in the weak coupling regime, it reproduces the Hartree approximation. However it is not limited to the small $U$ regime, and we make use of its versatility to perform calculations for arbitrary parameter values in the thermodynamical limit. 

All necessary details of the technique and numerical results can be found in the previous studies on the homogeneous 3D and 2D systems (specifically, for the evaluation of the compressibility and Landau parameters, in Ref.~\onlinecite{Lhou15,Fre16}).

\subsection{\label{sec:Hartree}Hartree-Fock evaluation}

We will compare results of the filling dependent compressibility for strong correlations in the electronic subsystems to those in the weak coupling case. For this comparison to be reasonable, the weak coupling regime must be realized within the same approach. While non-local interactions reduce to their Hartree approximation in the large D limit, it is not obvious in the 2D case that the slave-boson technique allows to gain the weak coupling results, in particular, when a non-local (attractive) interaction is included. For this purpose we evaluate the free energy of the 2D electronic system within standard Hartree-Fock perturbation theory and compare the results to those of the slave-boson evaluation with weak interaction. The extended 2D Hubbard model, which is expected to present the physics of the metallic electronic systems, is:

\begin{equation}
H = \sum_{\langle i,j\rangle,\sigma}t_{ij}^{\phantom{\dagger}}c_{i\sigma}^{\dagger}c_{j\sigma}^{\phantom{\dagger}} +U\sum_{i}\hat{n}_{i\uparrow}^{\phantom{\dagger}} \hat{n}_{i\downarrow}^{\phantom{\dagger}} + V\sum_{\langle i,j\rangle}^{\phantom{\dagger}}\hat{n}_{i}^{\phantom{\dagger}} \hat{n}_{j}^{\phantom{\dagger}}, 
\label{eq:model}
\end{equation}
In our evaluations below, we restrict the matrix elements $t_{ij}$ to $t$ for $\langle i,j\rangle$ a pair of nearest-neighbor sites and to $t'$ for next-nearest neighbor pairs on a square lattice; each pair is counted once. We compare the filling-dependent free energies in Hartree-Fock perturbation theory and in KR slave-boson evaluation~\cite{Fre16}.  

\begin{figure}[t]
\vskip0.5cm
\hskip-0.35cm\includegraphics[width=1.0\columnwidth]{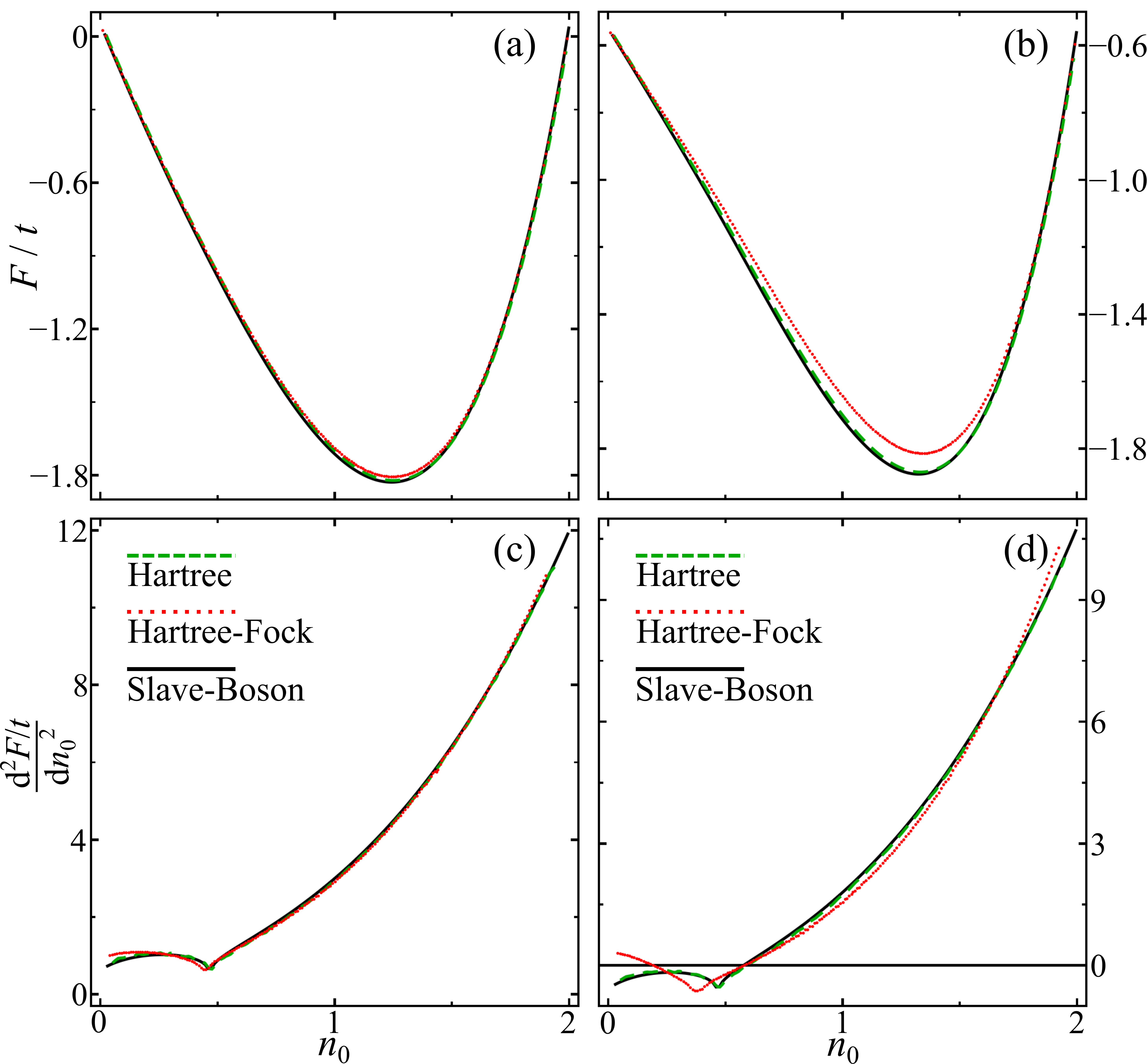}
\vskip0.3cm
\caption
{(Color online) Comparison of free energies in slave boson evaluation (black) and conventional perturbation theory within Hartree (dashed green line) and Hartree-Fock (dotted red line) approximation for $U=t$, $t'=-0.45t$ and (a) $V=-0.1\,t$ and (b) $V=-0.4\,t$. Panels (c) and (d)  display the second derivatives of the free energy with respect to the electron density $n$ for the respective nearest-neighbor interaction. The cusp is observed at the electron density where the van Hove singularity is positioned at the Fermi energy.}
\label{FreeEnergies}
\end{figure}

We take the Fourier transform of Hamiltonian (\ref{eq:model}) and make use of the commutation relations for fermionic operators:
\begin{align}
H = H_{\rm t}+H_{\rm U}+H_{\rm V}
\end{align}
with
\begin{align}
H_{t}&=-\sum_{\sigma,{\bf k}}\big\{2t\left[\cos(ak_{x})+\cos(ak_{y})\right]\notag\\
&\qquad\qquad+4t'\cos(ak_{x})\cos(ak_{y})\big\}c_{{\bf k}\sigma}^{\dagger}c_{{\bf k}\sigma}^{\phantom{\dagger}}\\
H_{U}&=\frac{U}{L^2}\sum_{{\bf k},{\bf p},{\bf q}}c_{{\bf k}+{\bf q}\uparrow}^{\dagger}c_{{\bf k}\uparrow}^{\phantom{\dagger}}c_{{\bf p}-{\bf q}\downarrow}^{\dagger}c_{{\bf p}\downarrow}^{\phantom{\dagger}}
\end{align}
\begin{align}
H_{V}&=\frac{V}{L^2}\sum_{\sigma,\sigma'}\sum_{{\bf k},{\bf p}}\bigg\{-\left[\cos{a(p_{x}-k_{x})}+\cos{a(p_{y}-k_{y})}\right]\times\notag\\
&\qquad\qquad \times c_{{\bf p}\sigma}^{\dagger}c_{{\bf p}\sigma'}^{\phantom{\dagger}}c_{{\bf k}\sigma'}^{\dagger}c_{{\bf k}\sigma}^{\phantom{\dagger}} +2c_{{\bf k}\sigma}^{\dagger}c_{{\bf k}\sigma}c_{{\bf p}\sigma'}^{\dagger}c_{{\bf p}\sigma'}\bigg\}.
\end{align}
Now the mean field approximation is performed
\begin{align}
H_{\rm U}^{\rm MFA}&=U\sum_{{\bf k},\sigma}n_{-\sigma}c_{{\bf k}\sigma}^{\dagger}c_{{\bf k}\sigma}^{\phantom{\dagger}}-UL^2n_\downarrow n_\uparrow\\
H_{\rm V}^{\rm MFA}&=V\sum_{{\bf k},\sigma}c_{{\bf k}\sigma}^{\dagger}c_{{\bf k}\sigma}^{\phantom{\dagger}}\bigg\{4(n_\uparrow+n_\downarrow)\label{eq:VHartreeFock}\\
&-\frac{2}{L^2}\sum_{{\bf p}\neq{\bf k}}\left[\cos{a(p_{x}-k_{x})}+\cos{a(p_{y}-k_{y})}\right]n_{{\bf p},\sigma}\bigg\}\notag\\
&+\frac{V}{L^2}\sum_{\sigma}\sum_{\substack{{\bf k},{\bf p}\\{\bf k}\neq{\bf p}}}\left[\cos{a(p_{x}-k_{\rm x})}+\cos{a(p_{y}-k_{y})}\right]\times\notag\\
&\qquad\qquad\times n_{{\bf p},\sigma}n_{{\bf k},\sigma} -2VL^2(n_\uparrow+n_\downarrow)^2\notag.
\end{align}
Here $L$ is the linear extension of the system, $a$ is the lattice spacing and $n_{\sigma}=1/L^2\sum_{\bf p}n_{{\bf p},\sigma}=1/L^2\sum_{\bf p}\langle c_{{\bf p}\sigma}^{\dagger}c_{{\bf p}\sigma}^{\phantom{\dagger}}\rangle$ the electron density with spin $\sigma$.
The first and the last term of Eq.~(\ref{eq:VHartreeFock}) constitute the Hartree term while the remaining two terms constitute the Fock term. The sign of the contribution of the Fock term to the energy is opposite to the sign of $V$. We take the parameter set $U=t$, $V=-0.1\,t$ or $-0.4\,t$, and $t'=-0.45\,t$.

From Fig.~\ref{FreeEnergies}, it is obvious that the Hartree approximation is not only consistent with the slave-boson result but is nearly identical in the weak coupling regime. 
This is a remarkable result as one expects that slave-boson theory is not appropriate for weak coupling and
small band filling. Yet, the KR scheme provides suitable projection factors for the kinetic energy and the
nearest-neighbor interaction so that its weak coupling limit is not only exact in the zero interaction case ($U=0$) but also
in first order of $U$. 

The exchange (Fock) correction for the non-local interaction $V$ is not included in the slave-boson evaluation. The free energies $F(n_0)$, as functions of the electron density $n_0 =\sum_\sigma n_\sigma$, differ pronouncedly close to their minima for sizable $V$---these minima are shifted from $n_0=1$ on account of the particle-hole symmetry breaking through a finite $t'$. However, for the compressibility we rather have to compare the second derivate of the free energies with respect to  $n_0$: the largest corrections from the Fock term arise for small densities (below the filling which displays the van Hove singularity related cusp). For long range Coulomb interaction, it is well known that  the Fock term dominates the direct Coulomb interaction only in the low density limit. This statement is also confirmed in our evaluation for a nearest-neighbor attractive interaction (see insets of Fig.~\ref{FreeEnergies}). Consequently, we argue that the weak coupling regime is attained in the slave boson approach even for finite but small interaction parameters---provided that the electronic density is not too low.
We test this assertion again in Appendix~\ref{appendix_B} where we compare the results for the capacitance of a heterostructure in Hartree-Fock and slave-boson evaluation.


\section{\label{sec:basics}Compressibility and Capacitance of a polar heterostructure}

\begin{figure*}[t]
\includegraphics[width=1.89\columnwidth]{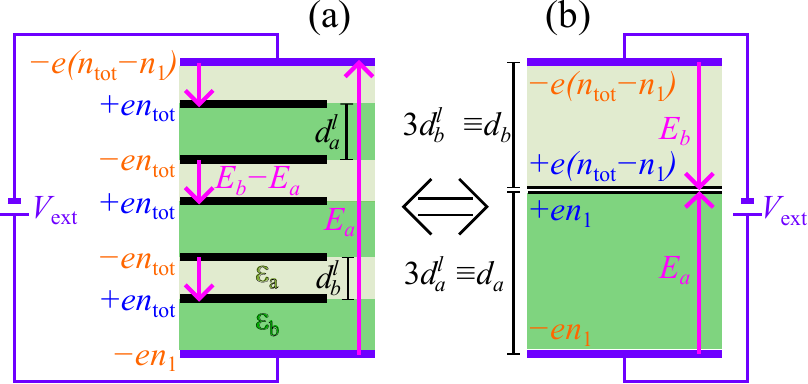}
\caption
{(Color online) (a) Polar heterostructure with a charge density of $\pm en_{\rm tot}$ on the polar layers and $-en_1$ on the interface and their corresponding electric fields (magenta). Electrons are transferred from the surface to the interface by an external voltage, which is included in $V_{\rm ext}$. (b) Equivalent circuit with the same electrostatic energy as (a). The distance between the electrodes and the layer of positive charge $d_a$/$d_b$ is given by the sum of distances between the polar layers $d_a^l$/$d_b^l$ of panel (a).}
\label{Setup}
\end{figure*}

\begin{figure*}[t]
\includegraphics[width=2.07\columnwidth]{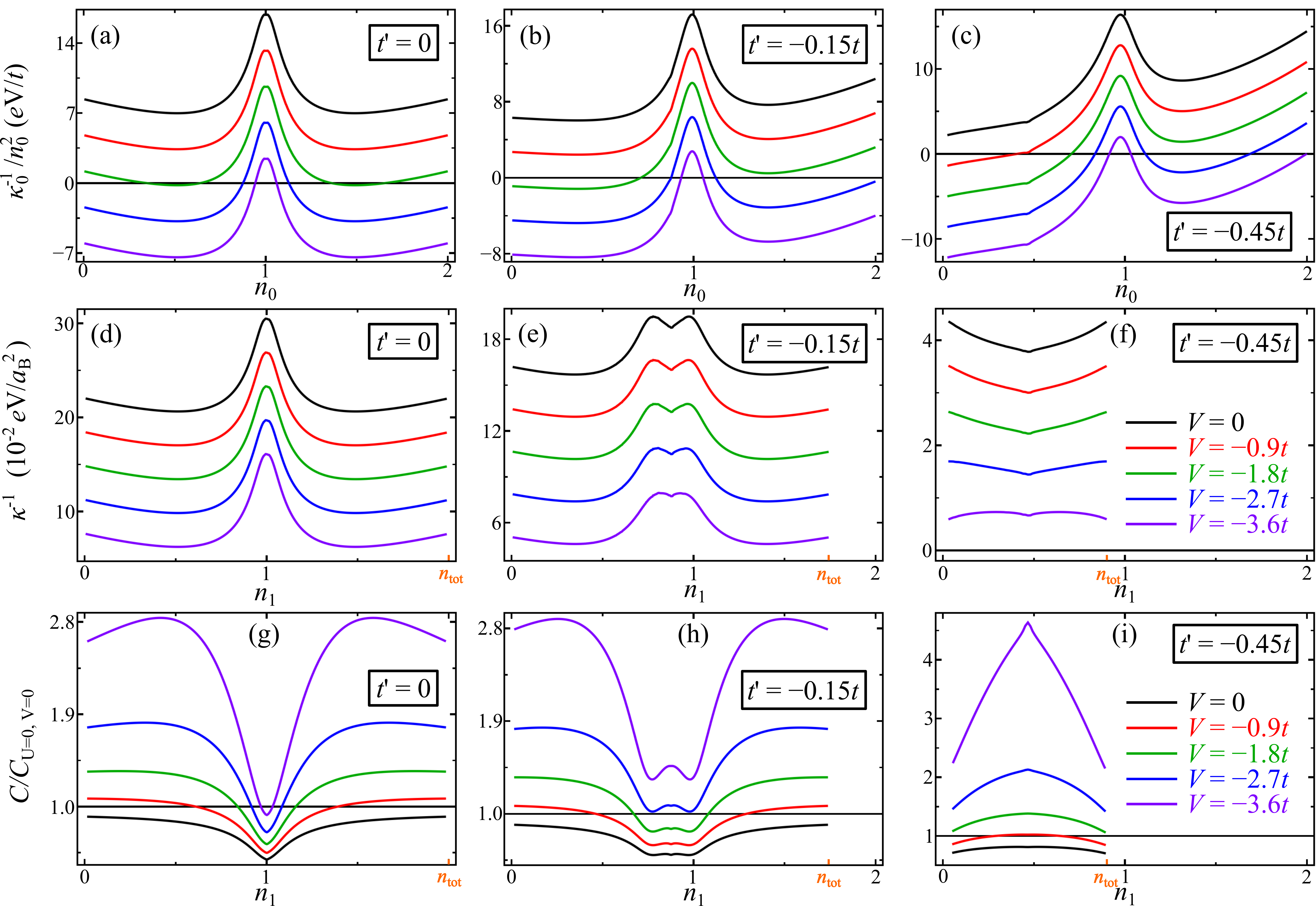}
\caption
{(Color online) (a)--(c) Inverse compressibilities of an isolated 2D strongly interacting electron system with  $U=9\,t$  
for three different values $t'$ (panels) and five different $V$ (see color code of (f) or (i)) as function of the electron density. (d)--(f) Inverse total compressibilities $\kappa^{-1}$ of a system composed of two identical interacting electron systems, as presented in (a)--(c), with  $t=0.5\,eV$, 
lattice constant $a=10\,a_{\rm B}$, and electrodes at a mutual effective distance $d/\varepsilon=4\,a_{\rm B}$. The inverse compressibilities
are displayed as functions of the electron density $n_1$ on the lower electrode. The electrostatic layout is illustrated in Fig.~\ref{Setup}. The total charge $n_{\rm tot}$ (orange) is chosen in such a way that at $n_1=n_{\rm tot}/2$ both systems have their Fermi energies at their respective van Hove singularities. (g)--(i) Differential capacitances of the above system normalized to the capacitance of two 2D electron systems with $U=0$ and $V=0$. The weak coupling results ($U=t$) are presented in Appendix~\ref{appendix_A} with Fig.~\ref{U1}.}
\label{U9}
\end{figure*}

We first consider a setup with two metallic 2D electron systems separated by a polar dielectric material. 
Such a configuration is realized, for instance, in gated LaAlO$_3$/SrTiO$_3$ heterostructures where a polar LaAlO$_3$
film is placed on a SrTiO$_3$ substrate. With a top gate on the LaAlO$_3$ film and with electronic reconstruction at the
interface between LaAlO$_3$ and SrTiO$_3$ into a (laterally) confined mobile electronic system, the heterostructure may be presented approximately as a multilayer system of the kind shown in Fig.~\ref{Setup} (left panel, the topmost plane is the gate and the lowest plane represents the interface). The charging of such a capacitance and the electronic density at the interface is tuned through the gate. The total mobile electronic density is $n_{\rm tot}$ and $n_1$ denotes the electron density which is
transferred from the top plane to the interface.
We investigated this type of heterostructure in a recent publication with focus on Rashba spin-orbit coupling at the interface~\cite{Ste15}. Here, we analyze the compressibility and capacitance of such a structure in dependence on electronic interaction parameters. We calculate the differential capacitance, which is the differentiation of the charge of a conducting plate with respect to the voltage, 
that is, the difference of the electrochemical potential with respect to the reference plate. Generally, the
capacitances of the conducting subsystems define a capacitance matrix but for the simple system comprised of two conducting 2D plates the matrix elements merge  into a single capacitance parameter $C$~\cite{Buettiker1993}.

The electronic compressibility is to be evaluated for the entire system---assuming that no subsystem decouples. Then the compressibility signals if the electronic system is prone to develop an instability towards a charge separated state. Such a 
phase separated state might be formed on a nanoscale with distinct properties, such as localization. However we will not investigate the various realizations of the symmetry broken state. We allow, right from the onset, a charge redistribution between mobile electrons on the surface electrode and the interface which is the electronic reconstruction driven by the polar dielectric in between the two metallic systems. 

In our model the total mobile electron charge density in the system, $-e\, n_{\rm tot}$,  is compensated by an equal amount of immobile positive background charge. The two electrodes (for example, surface plane and interface) are connected by an external voltage source, so that the difference in electrochemical potentials between the plates is given by $V_{\rm ext}$. A possible offset in chemical potentials between the two layers is included in this quantity. The electron density on the interface, $n_1$, and on the surface, $n_2=n_{\rm tot}-n_1$,  is tuned by applying the voltage $V_{\rm ext}$.  The free energy of the total system,
\begin{align}\label{eq:freeenergy}
F(n_{\rm tot},n_1)&=F_1(n_1)+F_2(n_{\rm tot}-n_1)+F_{\rm es}(n_{\rm tot},n_1)\notag\\
&\qquad\qquad-eV_{\rm ext}n_1A,
\end{align}
is  the sum of the free energy on the interface $F_1(n_1)$ and surface $F_2(n_{\rm tot}-n_1)$, an electrostatic contribution $F_{\rm es}(n_{\rm tot},n_1)$, and the term due to the difference in electrochemical potentials between the electrode plates. Here $A$ is the area of one plate and $e$ the elementary charge.

The free energy of the correlated electron system confined to the either plate, $F_{1,2}$, can be derived by the evaluation of Hamiltonian (\ref{eq:model})---either by slave-boson technique or within the Hartree (-Fock) approximation. 

The electrostatic energy for a polar heterostructure (Fig.~\ref{Setup}a), the equivalent circuit of which is depicted in Fig.~\ref{Setup}b, is:
\begin{align}\label{Fes}
F_{\rm es}(n_{\rm tot},n_1)&=\frac{2d_a\pi e^2A}{\varepsilon_a} n_1^2 +\frac{2d_b\pi e^2A}{\varepsilon_b}\left(n_{\rm tot}-n_1\right)^2
\end{align}
where $d_a$ ($d_b$) is the distance between the interface (surface) and the plane of positive charge in the equivalent circuit, while 
$\varepsilon_a$ ($\varepsilon_b$) is the corresponding dielectric constant. Here, we assume the configuration to be symmetric, i.e., $d_a/\varepsilon_a=d_b/\varepsilon_b\equiv d/(2\varepsilon)$. In LAO/STO a conducting interface develops for a distance $d$ of four unit cells, while the dielectric constant $\varepsilon$ is approximately~twenty. So we assumed that the effective distance between the plates is $d/\varepsilon=4a_{\rm B}$.

The total density $n_{\rm tot}$ is a thermodynamic variable, and the second derivative of the free energy with respect to $n_{\rm tot}$ determines the stability of the system.  In contrast, the density on the interface, $n_1$, is an internal variable that is fixed by the values of $n_{\rm tot}$ and $V_{\rm ext}$ and has to minimize the free energy of Eq.~(\ref{eq:freeenergy}):
\begin{align}\label{eq:minimizingcondition}
\frac{\partial F}{\partial n_1}&=\frac{\partial F_1}{\partial n_1}+\frac{\partial F_2}{\partial n_1}+\frac{\partial F_{\rm es}}{\partial n_1}-eV_{\rm ext}A =0
\end{align}
The differential capacitance is the differential change of the charge on one plate with applied voltage:
\begin{align}\label{eq:differentialcapacitance}
C_{\rm diff}^{-1}&=\frac{\partial V_{\rm ext}}{eA\partial n_1}\overset{(\ref{eq:minimizingcondition})}{=}\frac{1}{e^2A^2}\left(\frac{\partial^2 F_1}{\partial n_1^2}+\frac{\partial^2 F_2}{\partial n_1^2}+\frac{\partial^2 F_{\rm es}}{\partial n_1^2}\right)\notag\\
\end{align}
which implies that
\begin{align}\label{eq:differentialcapacitance2}
A/C_{\rm diff}&=\frac{1}{e^2A}\frac{\partial^2 F}{\partial n_1^2}
\end{align}
The last term in Eq.~(\ref{eq:differentialcapacitance}) constitutes the classical geometric capacitance. If the sum of the partial derivatives of $F_1$ and $F_2$ is negative, the differential capacitance is larger than the geometric capacitance. Note that for an isolated system, $\partial^2 F_i/\partial n_i$ corresponds to an inverse compressibility, and a negative value would signify instability. But here the interface and surface are coupled by the electric field and voltage source, so that the stability of the system is determined by the total derivative of the total free energy with respect to $n_{\rm tot}$:
\begin{align}\label{eq:compressibility}
\kappa^{-1}&=n_{\rm tot}^2\frac{\D^2(F/A)}{\D n_{\rm tot}^2}
\end{align}
For isolated (sub-) systems the quantities compressibility and capacitance are (up to a constant factor) the same. However, for the considered coupled system they can differ quite substantially (see Appendix~\ref{appendix_C}):
\begin{align}\label{compressibilitychargetransfer}
\frac{1}{n_{\rm tot}^2}\kappa^{-1}&=\frac{e^2A}{C_{\rm diff}}\left(1-\frac{\partial n_1}{\partial n_{\rm tot}}\right)\frac{\partial n_1}{\partial n_{\rm tot}}+\mathcal{D}F_{\rm es}
\end{align}
where the differential operator $\mathcal{D}$ is defined through $\mathcal{D}\equiv \left(\partial_{n_{\rm tot}}^2+\partial_{n_1}\partial_{n_{\rm tot}}\right)$.
In the case of the considered polar layout (see Fig.~\ref{Setup}a and Fig.~\ref{Setup}b), $\mathcal{D}F_{\rm es}=0$, and the compressibility is negative if the ``charge-transfer function''  $\partial n_1/\partial n_{\rm tot}$ is smaller than 0 or larger than 1, assuming a positive differential capacitance. 
Eq.~(\ref{compressibilitychargetransfer}) represents a generalization of a relation with $\mathcal{D}F_{\rm es}=0$ that was derived in Ref.~\onlinecite{Ste15}. 
There, the polar layout was considered exclusively; here we will investigate more general layouts in Sec.~\ref{sec:layout}.

The weak coupling results for the compressibility and capacitance are discussed in Appendix~\ref{appendix_A}. Here, we focus on the strong coupling analysis (Fig.~\ref{U9}) for the heterostructure of Fig.~\ref{Setup}. For comparison, we provide the inverse compressibility of an isolated 2D planar electron system (see Ref.~\onlinecite{Fre16}) in the topmost line of panels. In this section we assume that the electron systems and their respective parameters are the same for both plates.

First, for $t'=0$, the density dependence of the compressibility of the heterostructure (panel (d)) is similar to that of the single 2D plane (panel (a)): strong correlations ($U=9\,t$) produce a significant decrease in compressibility close to half-filling. We characterize the peak structure in the inverse compressibility as ``correlation peak''. Here we took $n_{\rm tot} =2.0$ for the heterostructure. Consequently, $n_1=1.0$ signifies that the single bands of the two plates (interface and surface) are each half-filled, and the correlation peak has its maximum at this value of $n_1$. Other values of $n_1$ are obtained by applying a corresponding voltage. Whereas the compressibility of the isolated 2D system becomes negative away from half-filling for sufficiently attractive nearest-neighbor interaction, this is not the case for the heterostructure. The polar dielectric material in between the plates provides an additional electrostatic energy and thereby stabilizes the system. The density dependence of the capacitance of the heterostructure (panel (g)) reflects the density dependence of the compressibility. The slight decrease of the capacitance for small and large $n_1$  (either most of the charge at the interface or at the surface) is caused by the fact that the attractive interaction is less effective at the band edges.

A finite value of $t'$ moves the van Hove singularity (vHs) from the center of the band. We take a filling of $n_{\rm tot} =1.75$ for $t'=-0.15t$ so that for equal filling in the two plates, the corresponding Fermi energies are placed at the vHs. The proximity to the vHs for both plates for $n_1\simeq n_{\rm tot}/2$ makes the structure in the compressibility more pronounced (panel (e)). At $n_{\rm tot}/2$ the inverse compressibility $\kappa^{-1}$ displays a cusp on account of the vHs. For $n_1$ larger than $n_{\rm tot}/2$, the interface approaches half-filling and the strong correlations lead to an increase in $\kappa^{-1}$, producing the correlation peak. Similarly, for $n_1$ smaller than $n_{\rm tot}/2$, the surface electron system approaches half-filling and 
$\kappa^{-1}$ increases. The capacitance has (correlation induced) minima at fillings where either plate is in an electronic state close to half-filling  (panel (h)). In the center we find a peak in the capacitance that originates from the attractive interaction $V$ augmented by the vHs which, for this filling, is positioned at the Fermi energies of both plates. This leaves a broad filling range where the capacitance is considerably enhanced with respect to its $U=0=V$ value. This happens when the filling in both plates is not close to half-filling.

Eventually, for  $t'=-0.45t$ and $n_{\rm tot} =0.93$ the correlation effects from the on-site repulsion $U$ are of minor importance because neither plate is close to half-filling. The nearest-neighbor interaction $V$ dominates the behavior of the compressibility and capacitance (see panels (f) and (i)). A strongly attractive $V$ increases the compressibility and capacitance. The strongest enhancement of the capacitance is at the van Hove filling (both plates have fillings so that the Fermi energies are placed at the vHs) at $n_1 = n_{\rm tot}/2$. At the largest and smallest value of $n_1$ one of the plates carries an electron system with filling close to $n_{\rm tot} =0.93$, and the on-site repulsion reduces the capacitance enhancement.

\begin{figure*}[t]
\includegraphics[width=2.0\columnwidth]{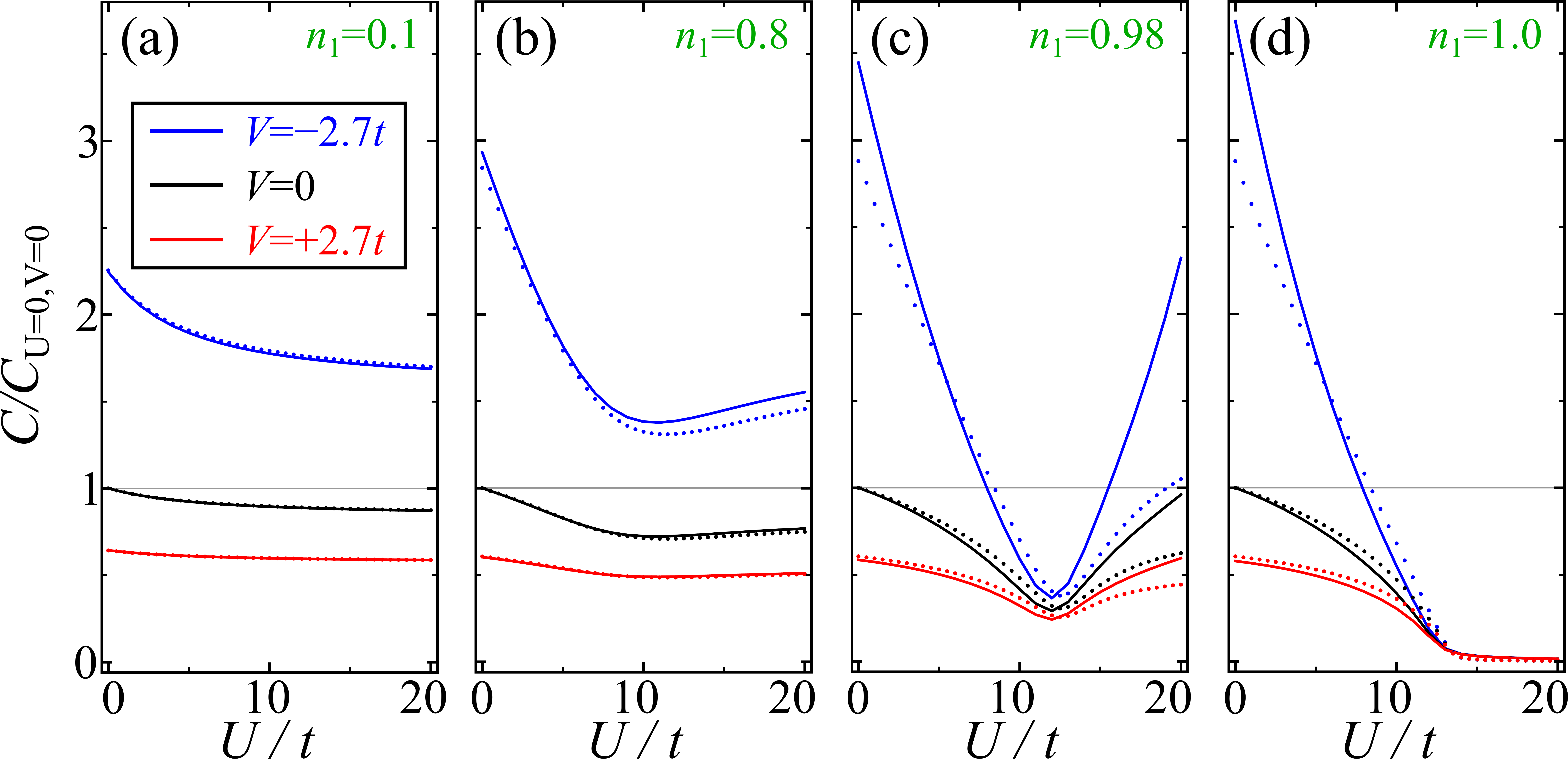}
\caption
{(Color online) Normalized capacitance in dependence on the on-site Coulomb interaction $U$ for different electronic densities. The mutual effective distance of the electrodes is $d/\varepsilon=4\,a_{\rm B}$. The nearest-neighbor interaction $V$ is tuned from positive to negative values. Next-nearest neighbor hopping is $t'=0$ (solid lines) and $t'=-0.45\,t$ (dotted lines). The voltage is adjusted so that the electronic density $n_1$ at the interface is $n_1=0.1$ (a), $n_1=0.8$ (b), $n_1=0.98$ (c), and $n_1=1.0$ (d). }
\label{CapStrongU}
\end{figure*}

For weak coupling---we present the results for $U=t$ in Appendix~\ref{appendix_A}---the correlation peak in the inverse compressibility at half-filling is absent. Instead one observes a dip in the inverse compressibility at the position of the van Hove singularity.


\section{\label{sec:largeU}Capacitance of strongly correlated electronic systems}

In Landau Fermi liquid theory, the electronic compressibility is controlled by the effective mass ratio $m^*/m_e$
and the Landau parameter $F_0^s$ 
(for a slave boson evaluation of $F_0^s$ and $m^*/m_e$ see Ref.~\cite{Fre16} for 2D, 
and for 3D see Ref.~\cite{Lhou15} and the review by Vollhardt~\cite{Vollhardt84} for the Gutzwiller approach to the Hubbard model). 
For fixed electronic densities on either metallic electrode, it is therefore expected that the capacitance decreases with increasing on-site interaction $U$: As $F_0^s$ is determined by $U$, the inverse compressibility 
increases correspondingly and the capacitance is suppressed. 
Indeed,  the capacitance decreases monotonically with $U$  for  $n_1=0.1$ at the interface and $n_2=n_{\rm tot}- n_1 = 1.9 $ at the surface (top) electrode (see Fig.~\ref{CapStrongU}a). In the following considerations, we keep the total density of electrons on the two metallic electrodes fixed to $n_{\rm tot}=2.0$. The electrostatic layout is still that of Fig.~\ref{Setup}. 

\begin{figure}[b]
\includegraphics[width=0.8\columnwidth]{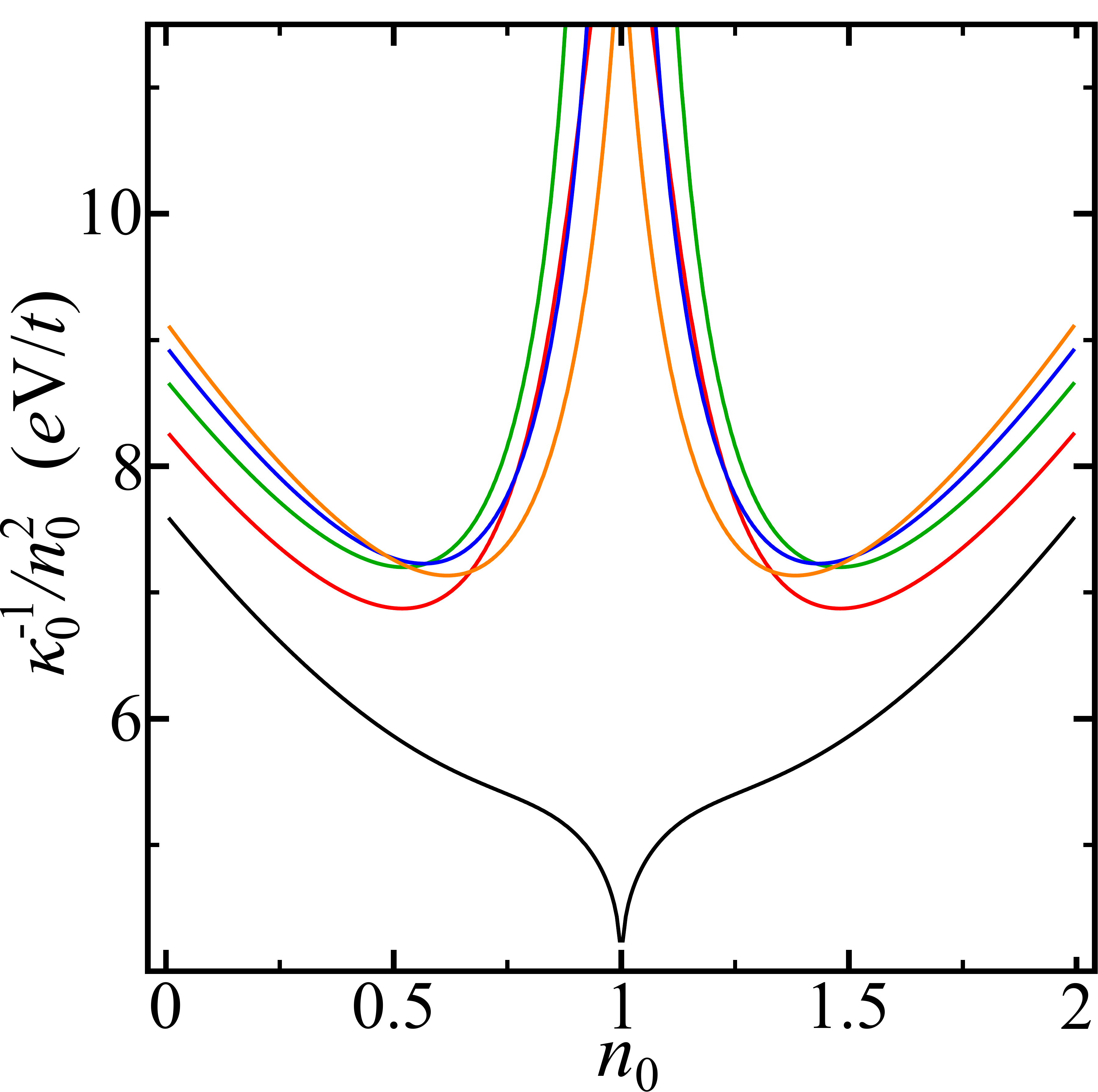}
\caption
{(Color online) Inverse compressibility of a 2D electron system as calculated from the second derivate of 
the free energy with respect to the electronic density $n_0$. 
The onsite interaction is $U=4\,t$ (black), $U=8\,t$ (red), $U=12\,t$ (green), $U=16\,t$ (blue), and $U=20\,t$ (orange). 
Here, $V=0$ and $t'=0$.}
\label{CompStrongU}
\end{figure}

Introducing a finite nearest-neighbor Coulomb interaction $V$ suppresses the capacitance further for repulsive interaction, albeit the dependence on $U$ is then rather weak (see the red line with $V=2.7\,t$ in Fig.~\ref{CapStrongU}a). An attractive interaction $V$ allows for an enhancement of the capacitance beyond its geometrical value for small $U$ 
(see the blue line with $V=-2.7\,t$ in Fig.~\ref{CapStrongU}a) and the decrease of $C$ with increasing $U$ is more distinct.

If the two electronic systems are each tuned to half-filling, that is $n_1=1.0 = n_2$, this decrease of $C$ with $U$ is even more pronounced and $C$ is zero beyond a critical value of $U_c$ where the electronic systems are in the incompressible Mott insulating state, independent of the value of $V$ (see Fig.~\ref{CapStrongU}d). Here, the transition from the Fermi liquid to a Mott insulating state is reached at $U_c\simeq 13\,t$. 

The striking observation is, however, that for intermediate densities the capacitance depends non-monotonically on $U$ (see Fig.~\ref{CapStrongU}b). We 
identify this behavior for densities in the range $0.6\lesssim n_1 < 1.0$ (and $n_2=2.0- n_1$) but it is most pronounced
close to half-filling ($n_1=0.98$, $n_2=1.02$ in Fig.~\ref{CapStrongU}c). In fact, $C(U)$ assumes a minimum at  $U_0\simeq 12\, t$ and increases for larger values of $U$. For positive values of $V$ the minimum persists but is less distinct than for attractive nearest-neighbor interaction. The $U_0$-values are above  $U_0\simeq 10\, t$ and approach $U_c$ in the vicinity of half-filling.

\begin{figure}[b]
\includegraphics[width=0.88\columnwidth]{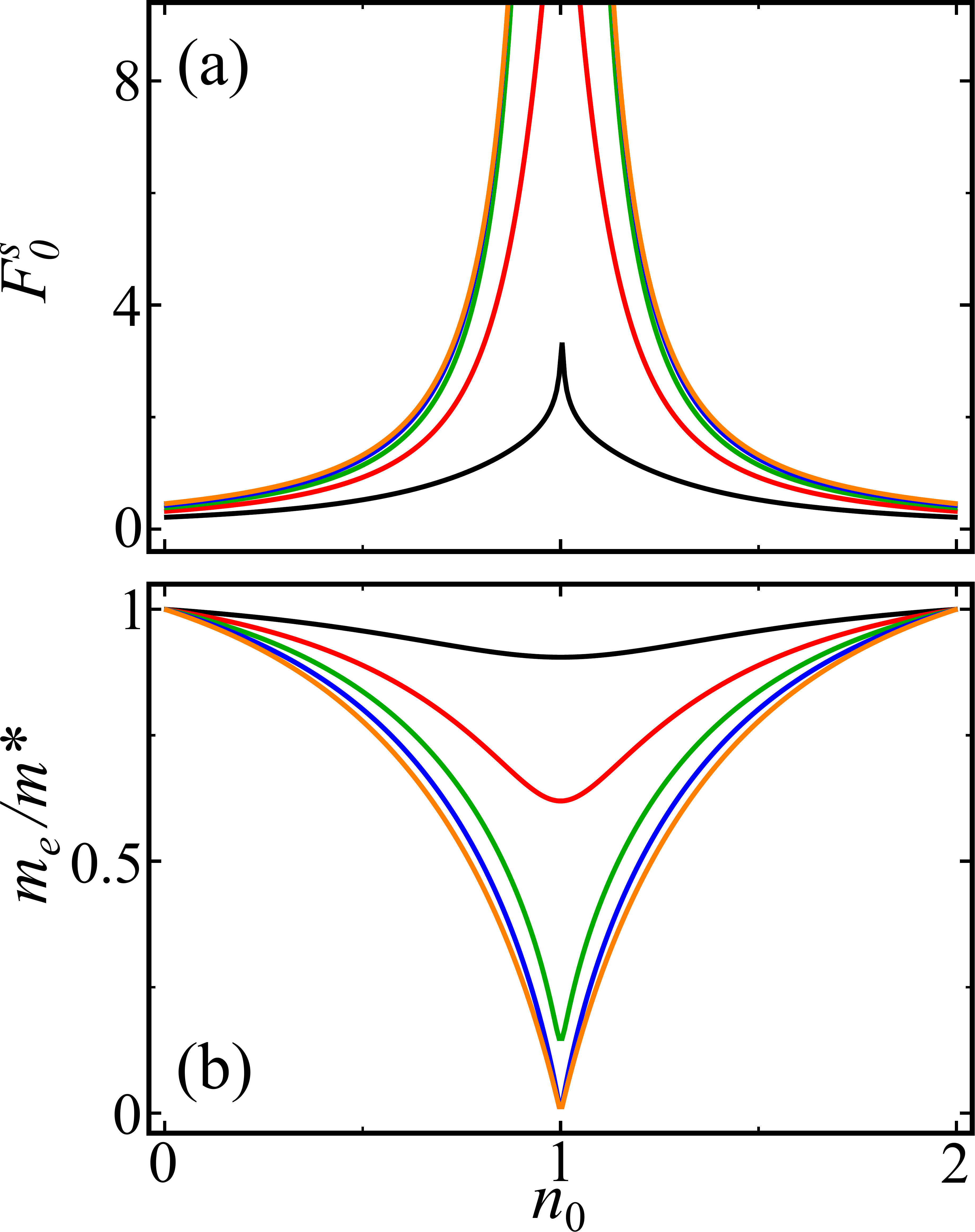}
\caption
{(Color online) Density dependence of the dimensionless Landau parameter $F_0^s$ (a) and the effective mass ratio $m_e/m^*$ (b)
for a 2D electron system. The onsite interaction is $U=4\,t$ (black), $U=8\,t$ (red), $U=12\,t$ (green), $U=16\,t$ (blue), and $U=20\,t$ (orange). Here, $V=0$ and $t'=0$.}
\label{Landau_n}
\end{figure}

\begin{figure}[b]
\includegraphics[width=0.88\columnwidth]{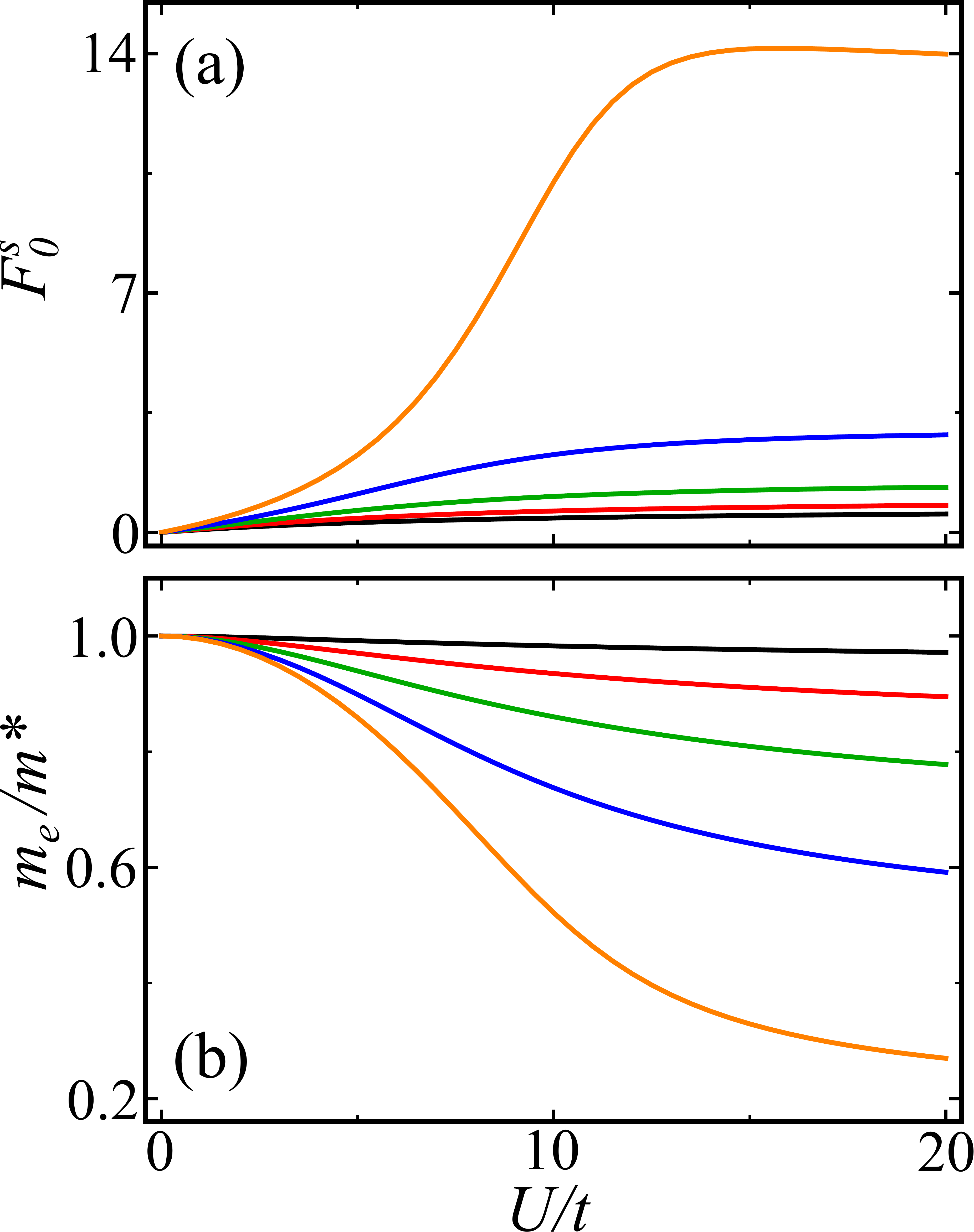}
\caption
{(Color online) $U$-dependence of the dimensionless Landau parameter $F_0^s$ (a) and the effective mass ratio $m_e/m^*$ (b)
for a 2D electron system. The electronic densities are $n_0=0.1$ (black), $n_0=0.3$ (red), $n_0=0.5$ (green), $n_0=0.7$ (blue), and $n_0=0.9$ (orange).}
\label{Landau_U}
\end{figure}

Inspection of the density dependent compressibility of a single 2D plane allows to draw conclusions on the origin of the non-monotonic behavior of $C(U)$ (see Fig.~\ref{CompStrongU}). We determine the inverse compressibility of the single plane directly from the second derivative of the free energy with respect to its electronic density $n_0$. 
The black curve ($U=4\,t$) corresponds to a weakly correlated 2D electron system, and a dip in the inverse compressibility is formed at the position of the van Hove singularity (cf.\ Fig.~\ref{U1}a 
in Appendix~\ref{appendix_A}). All other curves display a ``correlation peak'' in the inverse compressibility centered at half-filling (cf.\ Fig.~\ref{U9}a). At low densities and symmetrically at high densities (for $t'=0$) the inverse compressibility is larger, 
the higher the value of $U$ is. This holds for weak coupling as well as for strong coupling. 
However, at densities closer to half-filling the curves intersect in the strong coupling regime for $U\gtrsim 10\,t$. This signifies a
narrowing of the correlation peak and eventually leads to an enhancement of the capacitance with increasing $U$. The curve with 
$U= 8\,t$ is at intermediate coupling and crosses the curves with larger $U$ twice.

The observed narrowing of the correlation peak is caused by the interplay of two factors in the inverse compressibility: 
in Fermi liquid theory the inverse compressibility is controlled by the Landau parameter $F_0^s$ and the effective 
mass ratio $m^*/m_e$ according to
\begin{equation}\label{LandauComp}
\frac{\kappa_{0}(U=0)}{\kappa_{0}(U)}= (1+F_0^s)\cdot (m_e/m^*) 
\end{equation}
The dimensionless Landau parameter $F_0^s$ is obtained from the evaluation of the charge susceptibility (see Ref.~\cite{Fre16} 
for the 2D system); $F_0^s(U)$ diverges at half-filling for $U\rightarrow U_c$: $1+F_0^s(U)=(1-U/U_c)^{-2}$.
The second factor in~Eq.~(\ref{LandauComp}) is the inverse effective mass ratio $m_e/m^* =z_0^2$ which is
identified as the quasiparticle residue $z_0^2$ that approaches zero at half-filling for 
$U\rightarrow U_c$: $z_0^2=1-(U/U_c)^{2}$ \cite{Fre16,Lhou15,Vollhardt84}.
At half-filling, the inverse susceptibility diverges for $U\rightarrow U_c$.

Away from half-filling the divergence is lifted and the compressibility is non-zero also for $U$ above $U_c$. 
An increase of $1/\kappa_0(U)$ with the Landau parameter $F_0^s(U)$ is counteracted by a decrease with $m_e/m^*(U)$ (cf.\ Figs.~\ref{Landau_n}
(a) and (b)). For small $n_0\lesssim 0.5$ and also for large $n_0\gtrsim 1.5$, the $U$-dependence of the Landau parameter $F_0^s(U)$ dominates the product in Eq.~(\ref{LandauComp}), however close to half-filling the effective mass enhancement leads to a reduction of $1/\kappa_0(U)$ with increasing $U$. In fact, $F_0^s(U)$ starts to saturate beyond $U\simeq U_0$ (Fig.~\ref{Landau_U}a) whereas $m/m^*$ decreases further beyond $U\simeq U_0$ albeit at a reduced rate (Fig.~\ref{Landau_U}b).

With this decomposition of the inverse compressibility into a Landau quasiparticle interaction term and an (inverse) effective mass term, the result that the capacitance increases with $U$ beyond a critical value is plausible. 
 When the next-nearest neighbor hopping $t'$ is finite, the van Hove singularity is shifted away from half-filling (for $t' =-0.45\,t$ to a density
 of  approximately $0.47$). Yet, the minimum in $C(U)$ is still observable for densities $n_1$ and $n_2 =n_{\rm tot}-n_1$ around half-filling (see the
 dotted lines in Figs.~\ref{CapStrongU}b and \ref{CapStrongU}c). This observation illustrates that the non-monotonic behavior of $C(U)$ is a pure strong coupling effect.


\section{\label{sec:layout}Compressibility and stability in distinct layouts}

Even though the 2D electronic system may display a negative compressibility and consequently a tendency towards phase separation (see, e.g., Fig.~\ref{U9}(a)--(c)), the entire heterostructure, which is composed of one or two of these 2D layers, is not necessarily instable towards charge separation (see, e.g., Fig.~\ref{U9}(d)--(f)). It is the electrostatics that stabilizes the homogeneous state for the considered setup of Fig.~\ref{Setup}. The question arises if other layouts---with the same mobile electronic systems, $F_{1}$ and $F_{2}$, and the same distance $d$ between the electrodes---are more prone to phase separation.
It is evident that a different distribution of the positive background charge alters the electrostatic terms and thereby the total compressibility.

\begin{figure}[b]
\includegraphics[width=0.6\columnwidth]{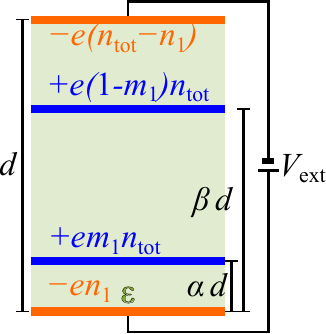}
\caption
{(Color online) Generalized electrostatic layout with fixed effective distance $d/\varepsilon$ between the planes of the electrons (orange). The positions of the two planes of positive background charge (blue) are parameterized by $\alpha$ and $\beta$ and the fractions of the total charge residing on them by $0\leq m_1\leq 1$.}
\label{GeneralizedSetup}
\end{figure}

\begin{figure*}[t]
\includegraphics[width=2.06\columnwidth]{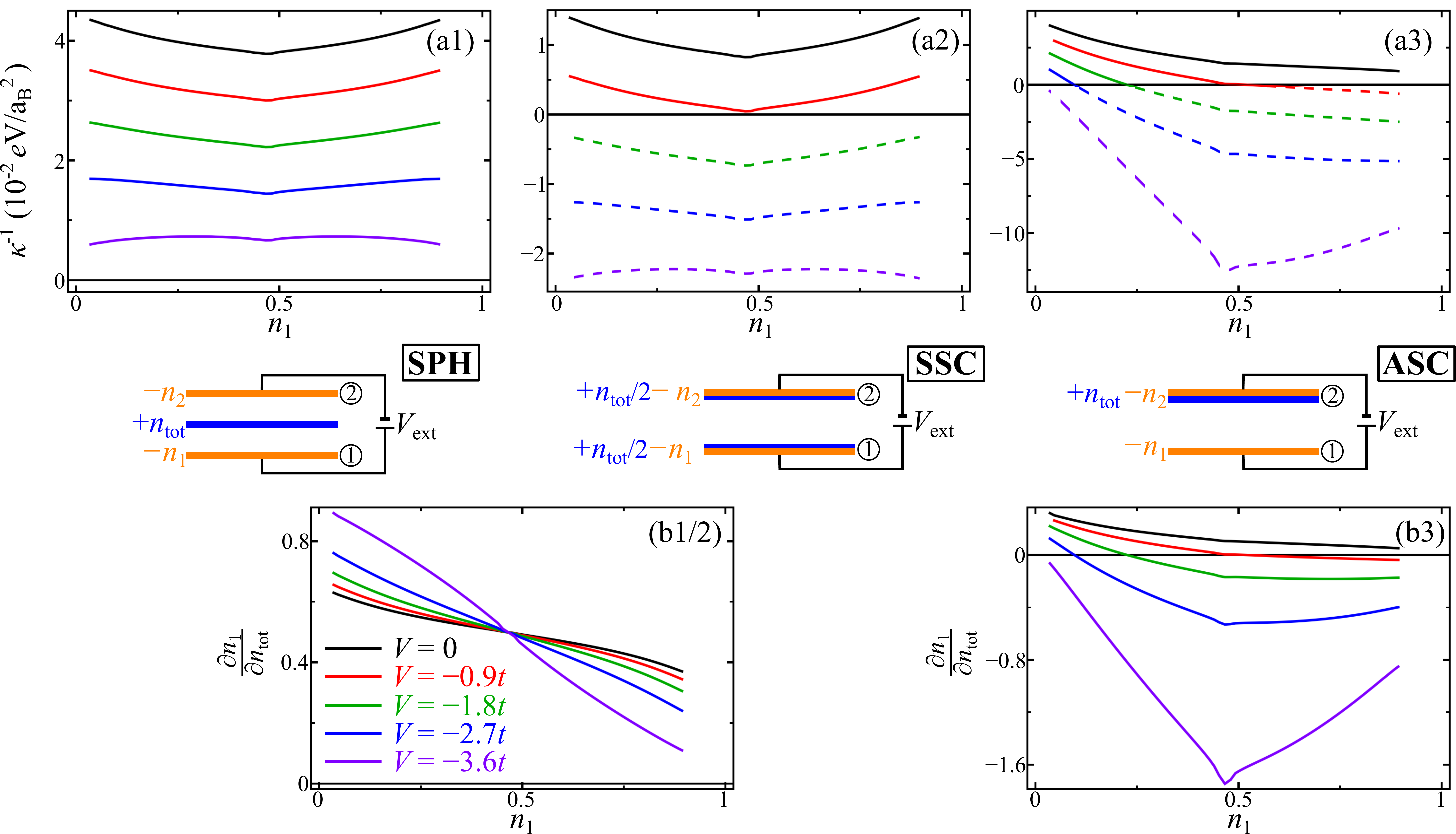}
\caption
{(Color online) Inverse compressibility (upper panels) and charge-transfer function (lower panels) for different distributions of the positive background charge (middle panels). The interacting electron systems on both electrodes are parameterized by $t=0.5\,eV$, $U=9\,t$, $t'=-0.45\,t$, $a=10\,a_{\rm B}$ and $d/\varepsilon=4\,a_{\rm B}$. The upper (lower) electrode comprises the electron system labeled 1 (2). For the three configurations $n_2= n_{\rm tot}-n_1$ holds, and all three capacitors have equal capacitances---but unequal compressibilities.}
\label{U9Configs}
\end{figure*}

In order to resolve this issue we introduce a generalized effective layout of positive background charge, 
characterized by the set of three parameters $\{\alpha, \beta, m_1 \}$
(c.f.~Fig.~\ref{GeneralizedSetup}). The distance $d$ between the two electrodes, which comprise the electron systems, is kept fixed. Two 2D layers with positive charge densities $+m_1n_{\rm tot}e$ and $+(1-m_1)n_{\rm tot}e$ (where $0\leq m_1\leq 1$) are located at a distance $\alpha d$ and $\beta d$ (where $0\leq\alpha\leq\beta\leq 1$) from the bottom electrode. This effective model covers a variety of physical configurations, inter alia:

The specific choice $\alpha=0$ and $\beta=1$ refers to the standard two-plate capacitor, where the positive charge populates the same layers as the electrons. The parameter $m_1$ allows for tuning the allocation of the positive charge between the plates.

For $\alpha=\beta$, all positive background charge of the effective model is placed in one plane and the setup is that of the polar heterostructures (c.f.~Fig.~\ref{Setup}\,(a) and (b)). The distance between the positively charged (effective) plane and the top electrode can be tuned either by using polar materials with different distances between the atomic layers $d_{\rm a}^l$ and $d_{\rm b}^l$, or by adding an insulating film and a terminating electrode on top of the polar heterostructure. Note that $\alpha=\beta=1$ (or $\alpha=\beta=0$) corresponds to the case where electrons from an electrode with neutralizing positive background ions are transferred into empty bands of a charge neutral electrode (by minimizing the free energy in the presence of a bias).

The electrostatic energy of this generalized effective layout is
\begin{align}\label{eq:generalelectrostaticen}
F_{\rm es}^g&=2D\big[\alpha n_1^2+(\beta-\alpha)\left(m_1 n_{\rm tot}-n_1\right)^2\notag\\
&\qquad +(1-\beta)\left(n_{\rm tot}-n_1\right)^2\big],
\end{align}
where
\begin{align}
D=\pi e^2 A d/\varepsilon=\frac{1}{4}e^2 A/C_{\rm geo}
\end{align}
is closely related to the inverse geometric capacitance $C_{\rm geo}$. With the free energy of Eq.~(\ref{eq:freeenergy}), we rewrite the second derivative with respect to the total electron density (see Appendix~\ref{appendix_C}):
\begin{align}\label{eq:secondderivativelayout}
\frac{\D^2F}{\D n_{\rm tot}^2}&=\frac{1}{F_1''+F_2''+4D}\bigg[\left(F_1''+F_2''+4D\right)
\left(F_2''+\frac{\partial^2F_{\rm es}^g}{\partial n_{\rm tot}^2}\right)\notag\\
&\qquad-\left(\frac{\partial^2F_{\rm es}^g}{\partial n_{\rm tot}\partial n_1}-F_2''\right)^2\bigg]
\end{align}
where $F_{1,2}'' \equiv \partial^2 F_{1,2}/\partial n_1^2$.
The inverse compressibility can be easily extracted from this quantity via Eq.~(\ref{eq:compressibility}). We emphasize that the capacitance is not affected by the distribution of the positive charge and is given (up to a factor of $e^2A^2$) by the fraction in front of the brackets. 

We now analyze three special configurations of the positive background charge: 
\begin{itemize}
\item the \textit{symmetric polar heterostructure} (SPH) with $\alpha=\beta=1/2$, as shown in Fig.~\ref{U9Configs}\,(a1)), 
\item the \textit{symmetric standard capacitor} (SSC) with $\alpha=0$, $\beta=1$, $m_1=1/2$, displayed in Fig.~\ref{U9Configs}\,(a2)),
\item the \textit{asymmetric standard capacitor} (ASC)  with $\alpha=\beta=1$, see Fig.~\ref{U9Configs}\,(a3)); the positive charge is on one electrode.
\end{itemize}
Corresponding setups of capacitors are displayed in Fig.~\ref{U9Configs}, jointly with the respective results for the inverse compressibility and the charge transfer function $\partial n_1/ \partial n_{\rm tot}$.

We can rewrite Eq.\,(\ref{eq:secondderivativelayout}),
\begin{align}\label{compressibilitysimplified}
A\kappa^{-1}_i&=n_{\rm tot}^2\frac{F_1''F_2''+D A_i}{F_1''+F_2''+4D}
\end{align}
with $i=\{{\rm SPH},{\rm SSC},{\rm ASC}\}$ and
\begin{align}\label{Alayouts}
A_{\rm SPH}&=2\left(F_1''+F_2''+2D\right)\notag\\
A_{\rm SSC}&=F_1''+F_2''\\
A_{\rm ASC}&=4F_2''\notag.
\end{align}

For nanoelectronics applications an enhancement of the capacitance beyond the geometrical value is often desirable (see the discussion in Refs.~\onlinecite{Kopp2009} and \onlinecite{Freericks2016}). A stable solution to $C> C_{\rm geo}$ requires that $F_1''+F_2''<0$, with a positive compressibility of the total system. However, the SSC system becomes instable for $F_1''+F_2''<0$ (see Appendix~\ref{appendix_C}, Eq.~(\ref{app:limit})). Consequently, an enhancement of the capacitance above the
geometrical value is not possible for the SSC layout (provided the electronic compressibility $\kappa_{SSC}$ stays positive). For $F_1''+F_2''>0$ the compressibility $\kappa_{\rm SSC}$ is always larger than the compressibility in the SPH system. This implies that the SPH is better suited to realize an enhanced capacitance in a thermodynamically stable state. In fact, the electrostatic energy in the SPH stabilizes the thermodynamic state.

The stability of the ASC layout depends on the sign of $F_2''$, i.e., it is best for a capacitance enhancement to pair the positive background layer with the electronic system, that is represented by a free energy with a positive second derivative. The electronic system of the second electrode should have a free energy with a negative second derivative in order to enhance the capacitance.

The strikingly different compressibilities of the three layouts with the same capacitance are shown in Fig.~\ref{U9Configs}\,(a1)--(a3). For this plot we took the same parameters for the two electron systems which were introduced in Sec.~\ref{sec:slaveboson}. The total density is adjusted in such a way, 
that both electron systems on the two electrodes are at the van Hove singularity for $n_1=n_2$. The density $n_1$ on the lower electrode is adjusted by an external voltage between the plates. In the thermodynamically unstable regime with negative compressibility, $\kappa^{-1}$ is represented by dashed lines. The SSC result is shifted to lower values of $\kappa^{-1}$ with respect to the SPH result; this observation is explained below. 

The relation between compressibility and capacitance is controlled by the charge-transfer function $\partial n_1/\partial n_{\rm tot}$ and by $\mathcal{D}F_{\rm es}$, as introduced in Eq.~(\ref{compressibilitychargetransfer}). In Appendix~\ref{appendix_C} we show that $\partial n_1/\partial n_{\rm tot}$ is in fact equal for the SPH and the SSC layout, so that the compressibilities of these two systems differ only by $\mathcal{D}F_{\rm es}$ (see Eq.~(\ref{compressibilitychargetransfer}). For the three model capacitors the $\mathcal{D}F_{\rm es}$ values are:
\begin{align}
\mathcal{D}F_{\rm es}^{\rm SPH}=0,\qquad
\mathcal{D}F_{\rm es}^{\rm SSC}=-D,\qquad \mathcal{D}F_{\rm es}^{\rm ASC}=0
\end{align}
This shift by $D$ of the inverse compressibility in the SSC (with respect to $\kappa^{-1}$ in the SPH) is visible in Fig.~\ref{U9Configs}---compare Fig.~\ref{U9Configs}\,(a1) and (a2). The ASC layout displays a negative inverse compressibility in the $V=-0.9\,t$ case  for densities larger than $n_{\rm tot}/2$ (see Fig.~\ref{U9Configs}\,(a3)). This results from $F_2''$ being negative for this filling (c.f. Fig.\,\ref{U9}\,(c) and Eq.\,(\ref{compressibilitysimplified}), (\ref{Alayouts})).

In systems with $\mathcal{D}F_{\rm es}=0$, the compressibility is positive for $0<\partial n_1/\partial n_{\rm tot}<1$, as is apparent from the comparison of 
Fig.~\ref{U9Configs}\,(b1) and Fig.~\ref{U9Configs}\,(b3) with the corresponding upper panels (a1) and (a3), respectively. 
For the SSC $\mathcal{D}F_{\rm es}= -\pi e^2 A d/\varepsilon$ and one verifies for $n_1=\frac{1}{2}n_{\rm tot}=n_2$ that $\partial n_1/\partial n_{\rm tot}=\frac{1}{2}$ (compare Fig.~\ref{U9Configs}~(b2)) and that, with Eq.~(\ref{compressibilitychargetransfer}), the well-known relation $\kappa^{-1}/n_1^2 + 4\pi e^2 d/\varepsilon = e^2 A/ C_{\rm diff}$ holds (see, for example, Ref.~\onlinecite{Kopp2009}).

\begin{figure}[t]
\includegraphics[width=0.99\columnwidth]{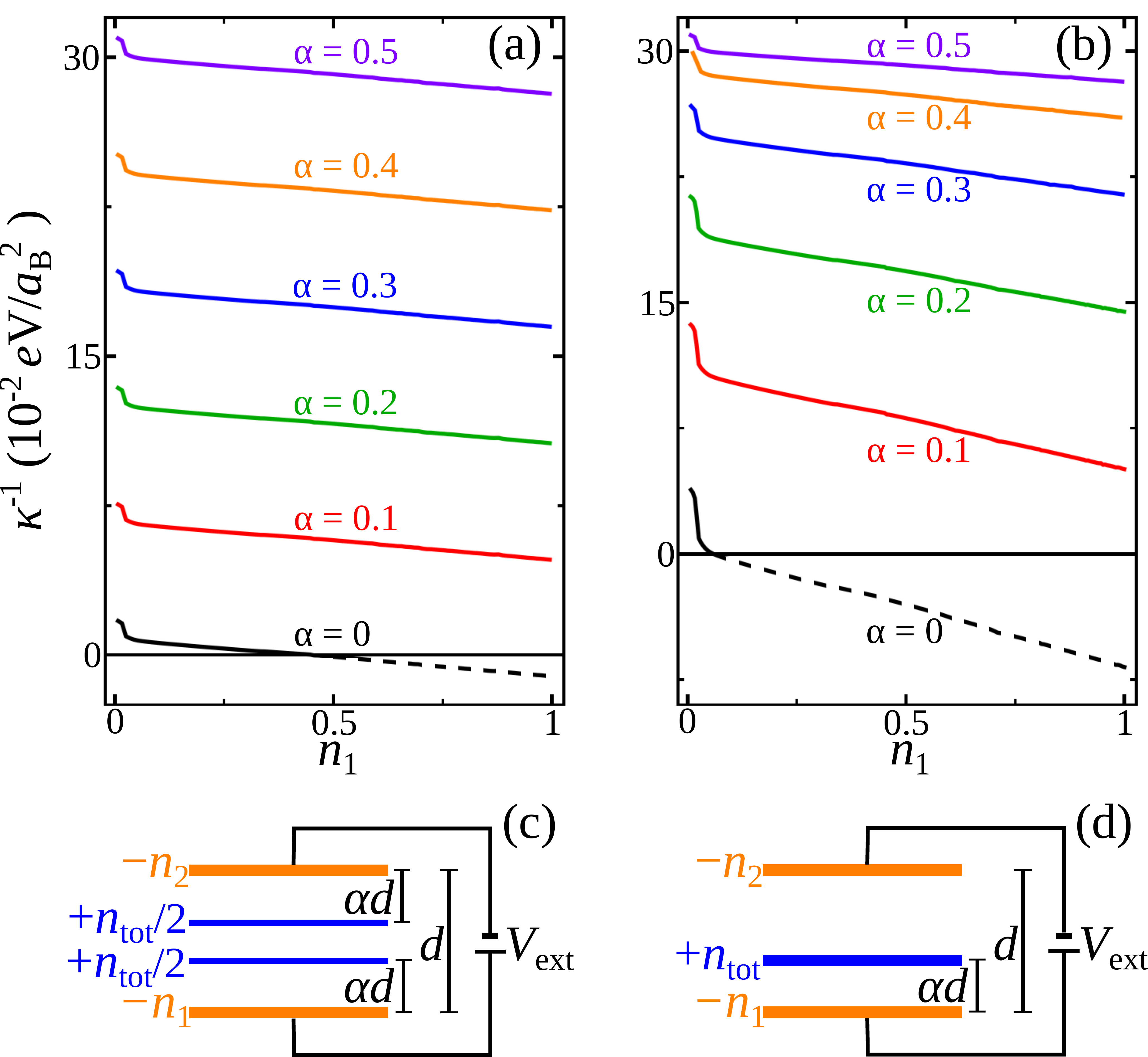}
\caption{(Color online) (a,b) Inverse compressibility of a system with Rashba spin-orbit coupling at the interface and a metallic surface for parameters as in Ref.~\onlinecite{Ste15} (see Fig.~6a therein). The corresponding configuration of the positive background charge is depicted in the panels below. For $\alpha=0.5$ the SPH-setup is recovered, which was used in Ref.~\onlinecite{Ste15}.}
\label{SOC}
\end{figure}

In experimental setups the top electrode is often a metallic electrode for which correlation effects can be neglected. Layouts with unequal electrodes are investigated in Appendix~\ref{appendix_B}. There the top electrode is chosen to be an
uncorrelated 2D metallic system and the bottom electrode a correlated 2D metal.

The dependence of the compressibility on the electrostatic layout is also relevant for heterostructures with electronic systems other than those represented by the extended Hubbard model: which configurations allow for a negative compressibility with phase separation? In Refs.~\onlinecite{Grilli2012,Bucheli2014,Seibold2015} an electron system at  the interface of LaAlO$_3$ and SrTiO$_3$ was suggested to display electronic phase separation on account of a finite Rashba spin-orbit coupling. In Ref.~\onlinecite{Ste15} this system was supplemented by electronic surface states, resembling the SPH, and the possibility of a state with negative compressibility was analyzed. For values of the spin-orbit coupling consistent with experiments, the compressibility of the system was found to be positive~\cite{Ste15}.

Here we investigate two realizations of the electrostatic layout which are variants of the general scheme in Fig.~\ref{GeneralizedSetup}: in the first, the total positive background charge is split in half and both charge fractions are moved from the center plane by the same distance towards the electrodes (Fig.\,\ref{SOC}\,(a) and (c)). The second variation keeps the background charge in plane (Fig.~\ref{SOC}\,(b) and (d)). The spin-orbit coupling and all electronic system parameters are fixed. 

For the first alternative the layout of the positive background charge is symmetric. Then the charge-transfer function is identical for all values of the distance parameter $\alpha$ (see Appendix~\ref{appendix_C}) and the compressibilities differ only by the (density independent) electrostatic term $\mathcal{D}F_{\rm es}(\alpha)$ (c.f.~Fig.~\ref{SOC}\,(a)). For $\alpha=0$ the SSC is recovered and we find that the inverse compressibility of the total system can become negative. In the case of the layout shown in Fig.~\ref{SOC}\,(d), a reduction of $\alpha$ merges the layout finally with the ASC layout (for $\alpha=0$) which may also acquire a negative electronic compressibility. 
We argue that the electrostatic energy in the polar heterostructure prevents a phase separation---the SPH is represented by the purple curves with $\alpha=0.5$ in Fig.~\ref{SOC}.  Other configurations, however, have a stronger tendency towards phase separation---such as the SSC (Fig.~\ref{SOC}\,(a) and (c) with 
$\alpha=0$) or the ASC (Fig.~\ref{SOC}\,(b) and (d) with $\alpha=0$).


\section{Conclusions}\label{sec:conclusion}

The electronic compressibility characterizes the electronic state of capacitive heterostructures through its dependence on the gate bias; this voltage controls the electronic density in the system. For an isolated 2D layer, the qualitative density dependence of the compressibility can be readily understood (compare Figs.~\ref{U9}(a)--(c) and Figs.~\ref{U1metalSBHFCap}(a)--(c) for the strongly and weakly interacting 2D systems, respectively): for weak coupling the compressibility represents the density of states and one observes a pronounced dip in $\kappa_0^{-1}$ for the density where the Fermi energy is tied to the van Hove singularity. For strong coupling, electronic correlations are responsible for the peak structure in $\kappa_0^{-1}$ in a sizable filling range around $n_0=1$ (``correlation peak''). For a negative nearest-neighbor interaction $V$, introduced as antagonist to the repulsive on-site interaction $U$, one observes a shift of $\kappa_0^{-1}(n_0)$ towards smaller or even negative values. 
It should be emphasized here that our results are not limited to the model with nearest-neighbor interaction. In fact, they apply to systems with arbitrary screened Coulomb interaction as well, since the latter enters $\kappa_0$ and the saddle-point equations only through its $k=0$ Fourier component \cite{Lhou15}. 
In our analysis of the (inverse) electronic compressibility of a 2D system we do not find an interplay of the on-site Coulomb repulsion $U$
and the nearest-neighbor interaction $V$. This might be due to the slave-boson saddle-point evaluation but it may well be that inhomogeneous 2D states have a more complex dependence on these interaction scales.  Such inhomogeneous states have not been considered in the present work. 

These findings are to be re-examined for a heterostructure with at least two coupled metallic 2D systems (see Figs.~\ref{U9}(d)--(f) for the strongly interacting 2D system with a polar film between the metallic plates, as in Fig.~\ref{Setup}). 
The total compressibility is always positive provided that the compressibilities of the subsystems are positive.
One result is of particular importance: The total electronic compressibility $\kappa$ of the heterostructure 
can stay positive even if the compressibilities of the metallic subsystems are negative, that is the heterostructure is more stable with respect to the formation of a phase-separated electronic state. It is the interlayer electrostatic term which, apart from the intralayer electronic interaction, influences the electronic reconstruction in the heterostructure and keeps the total compressibility positive.

There is no general recipe to identify systems with negative or positive compressibility from the outset. Each system has to be 
evaluated self-consistently with respect to its charge distribution, depending on coupling parameters such as $U$, $V$ or the spin-orbit coupling
and the layout of the heterostructure.
We investigated different layouts and identified the symmetric polar heterostructure (SPH setup in Fig.~\ref{U9Configs}) as the configuration in a large  class of systems (continuously characterized by three parameters) which is most robust (smallest positive $\kappa$) with respect to other configurations with the same capacitance. A system that is more susceptible to phase separation is the standard capacitor with two electrodes and no polar dielectric, provided that an electronic coupling such as an attractive nearest-neighbor interaction or a Rashba spin-orbit coupling (see Figs.~\ref{U9Configs} and \ref{SOC}) drives the metallic plates into a negative compressibility state. The standard capacitor always displays a negative total compressibility if the two electrodes are in a negative compressibility state, that is, it is thermodynamically instable if no further terms (e.g.\ from the lattice) keep the total compressibility positive. 

The capacitance is distinct from the compressibility of the heterostructure. It is the response of the charge density of a plate to a voltage difference applied to the two electrodes---whereas the compressibility is the response of the total charge density to the chemical potential of the electronic system. In our model system, the capacitance can achieve a large enhancement with respect to its geometric value. In particular, this is realized  for sizable $t'$ so that the van Hove singularity, which may drive the enhancement for attractive $V$, is moved from the band center. 
Otherwise, a repulsive
$V$ reduces the capacitance, and a repulsive on-site interaction $U$ is very effective to suppress the capacitance when 
the electronic systems are close to half-filling. However, we also find that the capacitance can be enhanced when $U$ is larger than a critical value. 
In this regime of very strong coupling, the $U$-induced enhancement of the effective mass dominates the counteracting increase 
of the effective quasiparticle interaction parameterized by $F_0^s$ in the compressibility. 
Such an anomalous behavior of $C(U)$ is most pronounced close to half-filling.
In order to identify heterostructures with large capacitance one should in general avoid electronic systems with strong on-site interaction on the electrodes. However it is also necessary to work with a sufficiently stable system---and we find that  the heterostructures with polar dielectrics abide by this characterization.

\section*{Acknowledgments} 
We thank F.~Loder and J.~Mannhart for helpful discussions. The authors acknowledge the financial support of the DFG through the TRR~80, and of the French Agence Nationale de la Recherche (ANR), through the program ``Investissements d'Avenir'' (ANR-10-LABX-09-01), LabEx EMC3.

\appendix

\section{Weak coupling results}\label{appendix_A}

\begin{figure*}[t]
\includegraphics[width=2.07\columnwidth]{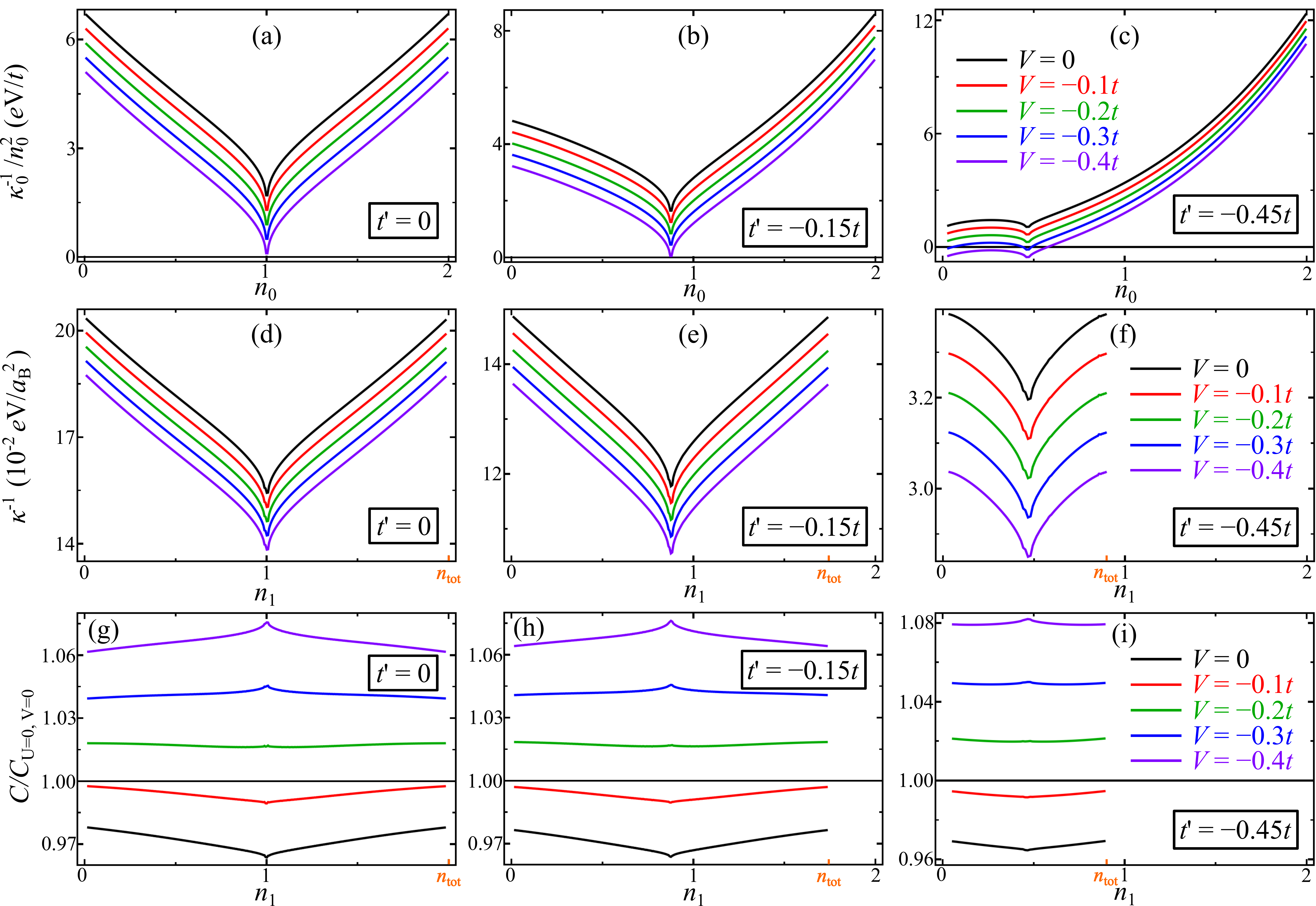}
\caption
{{(Color online) (a)--(c) Inverse compressibilities of an isolated 2D weakly interacting electron system with $U=t$ 
for three different values $t'$ (panels) and five different $V$ (see color code of (f) or (i)) as function of the electron density. (d)--(f) Inverse total compressibilities $\kappa^{-1}$ of a system composed of two identical interacting electron systems, as presented in (a)--(c), with  $t=0.5\,eV$, 
lattice constant $a=10\,a_{\rm B}$, and electrodes at a mutual effective distance $d/\varepsilon=4\,a_{\rm B}$. The inverse compressibilities
are displayed as functions of the electron density $n_1$ on the lower electrode. The electrostatic layout is illustrated in Fig.~\ref{Setup}. The total charge $n_{\rm tot}$ (orange) is chosen in such a way that at $n_1=n_{\rm tot}/2$ both systems have their Fermi energies at their respective van Hove singularities. (g)--(i) Differential capacitances of the above system normalized to the capacitance of two 2D electron systems with $U=0$ and $V=0$.}}
\label{U1}
\end{figure*}

We analyze the same setup as described in Fig.~\ref{U9}, here with $U=t$ instead of $U=9\,t$, that is, 
the electron systems are characterized by weak correlations. Since for smaller on-site repulsion it is also reasonable that the inter-site attraction is smaller, we used the same ratio of $U/V$ as in Fig.~\ref{U9}. As this means smaller values of $|V|$, the inverse compressibilities of the single isolated Hubbard systems shown in Fig.~\ref{U1} (a)--(c) are larger than in Fig.~\ref{U9}. Moreover, the inverse compressibilities do not display ``correlation peaks'', in marked contrast to the results for $U=9\,t$.
 
The absence of the correlation peak is reflected in the inverse compressibilities of the total system (Fig.~\ref{U9} (d)--(f)), the structure of which is controlled by the single layer compressibilities. Again, the chosen densities $n_{\rm tot}$ for different values of $t'$ ensure that at $n_1=n_{\rm tot}/2$ the Fermi energy of both systems is at the van Hove singularity. This is accomplished by $n_{\rm tot}=2.0$ for $t'=0$, $n_{\rm tot}=1.75$ for $t'=-0.15\,t$ and $n_{\rm tot}=0.93$ for $t'=-0.45\,t$. 
 
With no correlation peak in the inverse compressibility, the differential capacitance (Fig.~\ref{U9} (g)--(i)) does not display a dip at $n_1=n_{\rm tot}$. We normalized the capacitance to that of a $U=0$ and $V=0$ system. It is evident that for weak inter-site attraction (black and red curves) the finite on-site repulsion $U=t$ induces a differential capacitance smaller than $C_{U=0,V=0}$. Since $|V|$ is taken to be smaller, the isolated single layer inverse compressibilities Fig.~\ref{U1} (a)--(c) are larger (and mostly positive) and, hence, the achieved increase of the capacitance is smaller than in Fig.~\ref{U9} (g)--(i).

\section{Heterostructure with unequal electrodes}\label{appendix_B}

\begin{figure}[t]
\includegraphics[width=0.7\columnwidth]{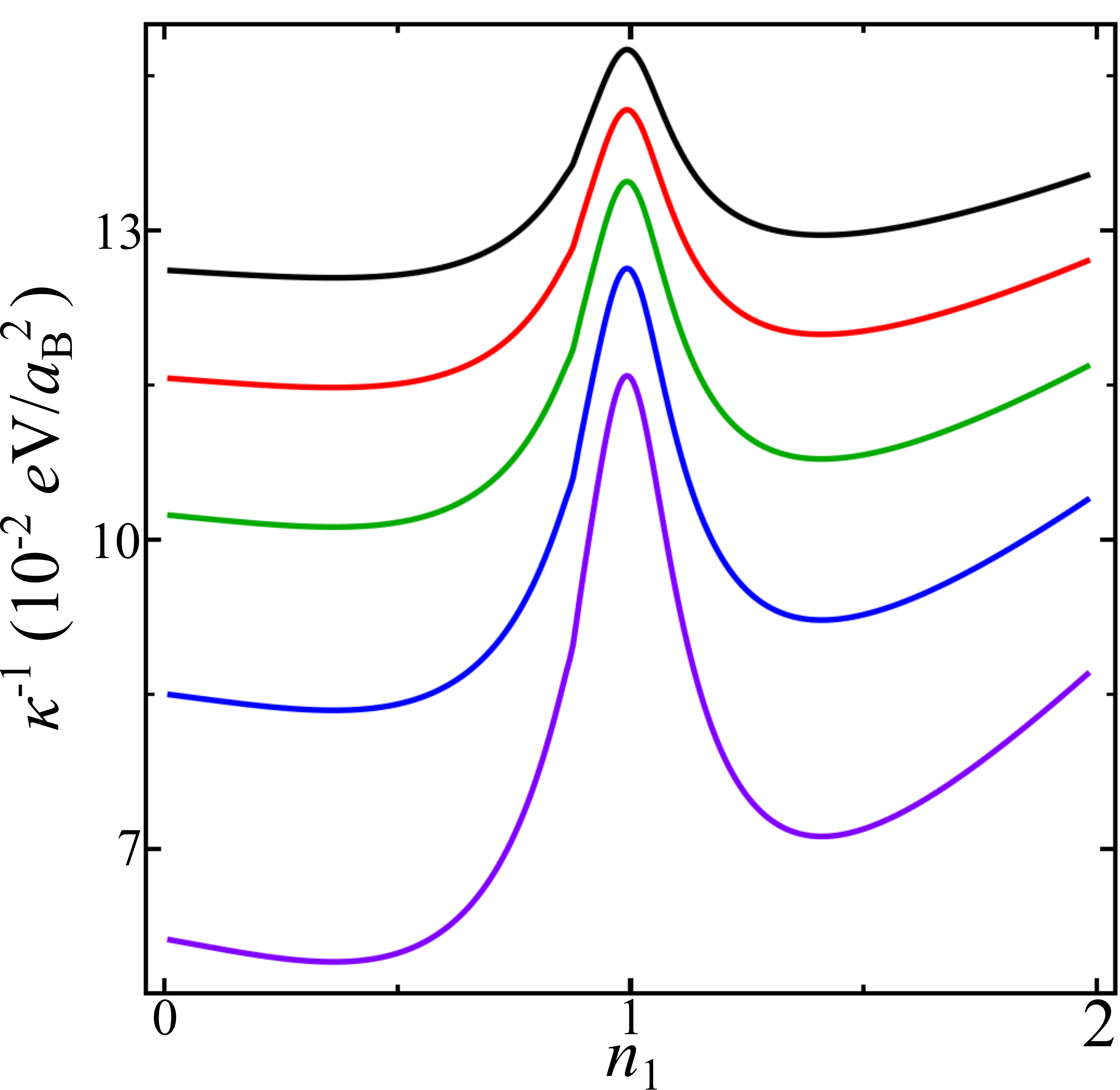}
\caption
{(Color online) The inverse compressibility when the surface electron system of Fig.~\ref{U9}\,(e) is replaced by a metal with $m_2/m_e=5$. The interface electron system is characterized by strong correlations, $U=9\,t$, and by $t'=-0.15\,t$. $\kappa^{-1}$ is smaller than that of Fig.~\ref{U9}\,(e). The color coding for the different values of $V$ is the same as in Fig.~\ref{U9}\,(f).}
\label{U9metalComp}
\end{figure}

\begin{figure}[b]
\includegraphics[width=0.7\columnwidth]{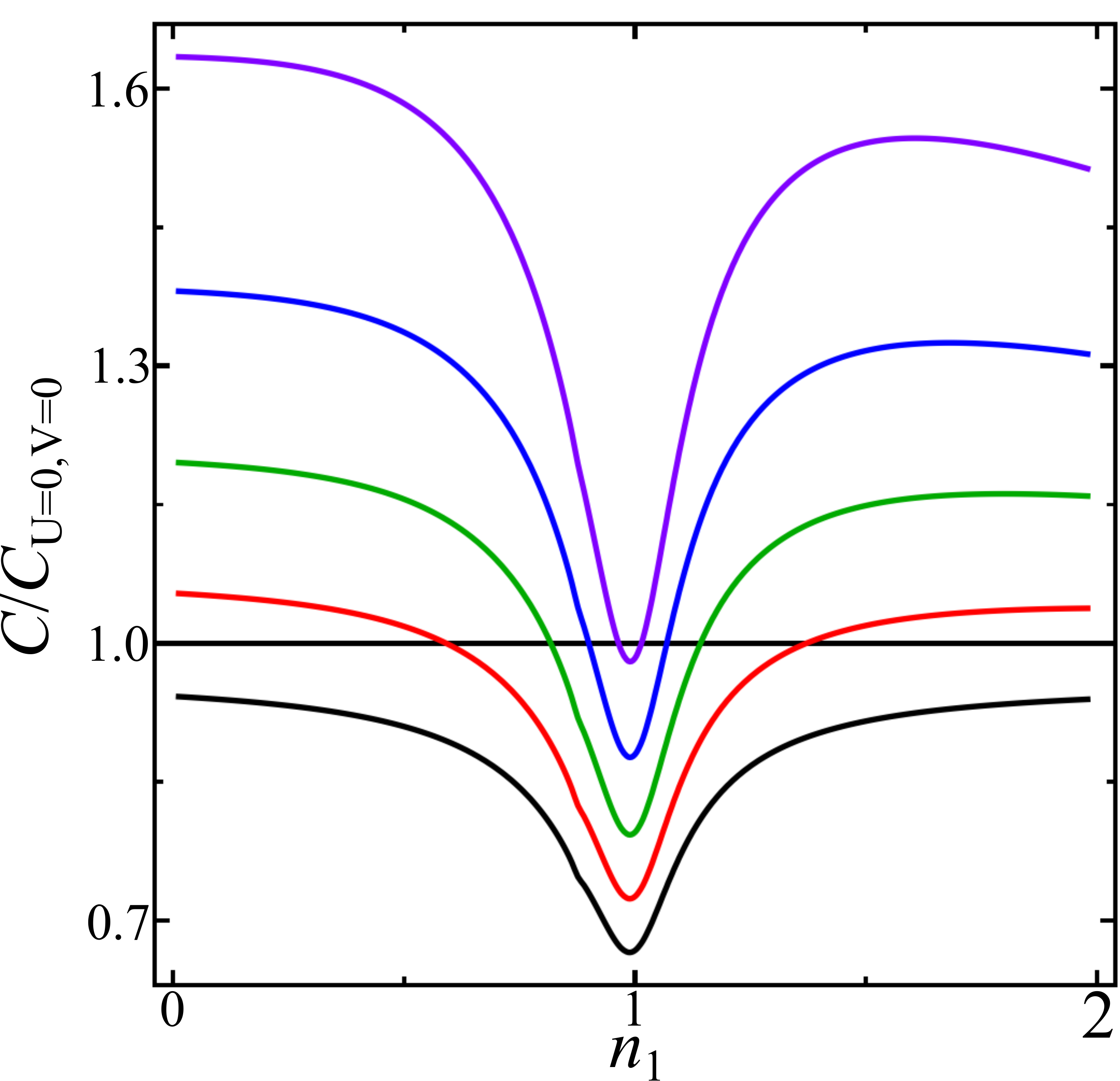}
\caption
{(Color online) The capacitance for the heterostructure of Fig.~\ref{U9metalComp} becomes less enhanced above the $U=0$, $V=0$ value as compared to the symmetric setup with both electrodes comprising a strongly correlated electron system.}
\label{U9metalCap}
\end{figure}

\begin{figure}[t]
\includegraphics[width=0.7\columnwidth]{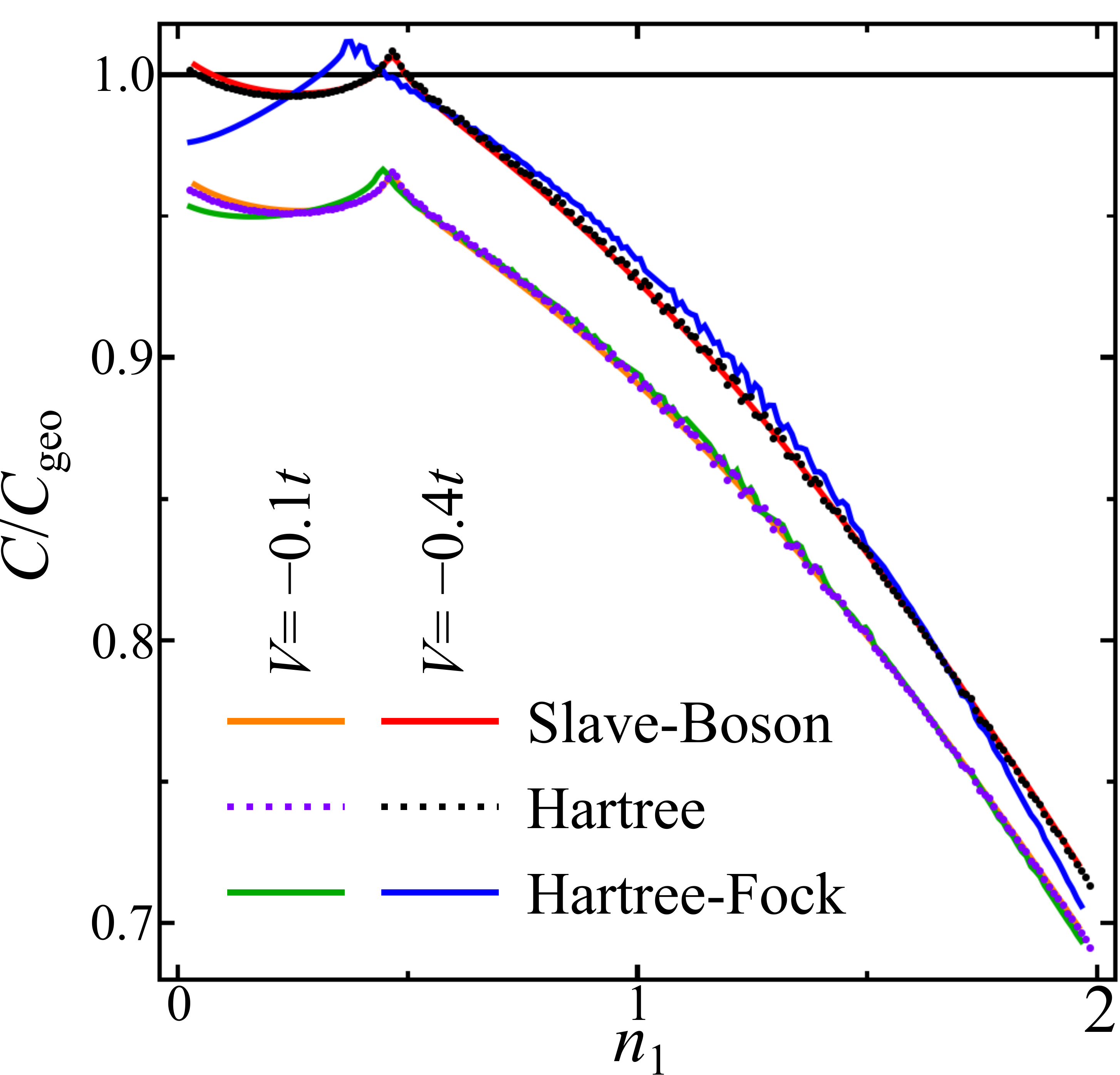}
\caption
{(Color online) The capacitance of a system with electrons on the lower electrode described by the extended Hubbard model for different approximations is normalized to the geometric capacitance $C_{\rm geo}=4\pi d/\varepsilon$. The mass of the metallic electrons on the surface is $m_2/m_e=5$ and the parameters of the interface electrons are $t=0.5\,eV$, $U=t$, $t'=-0.45\,t$ and $a=10\,a_{\rm B}$. The system is in the SPH configuration and the effective interplate distance is given by $d/\varepsilon = 4\,a_{\rm B}$.}
\label{U1metalSBHFCap}
\end{figure}

\begin{figure}[b]
\includegraphics[width=0.7\columnwidth]{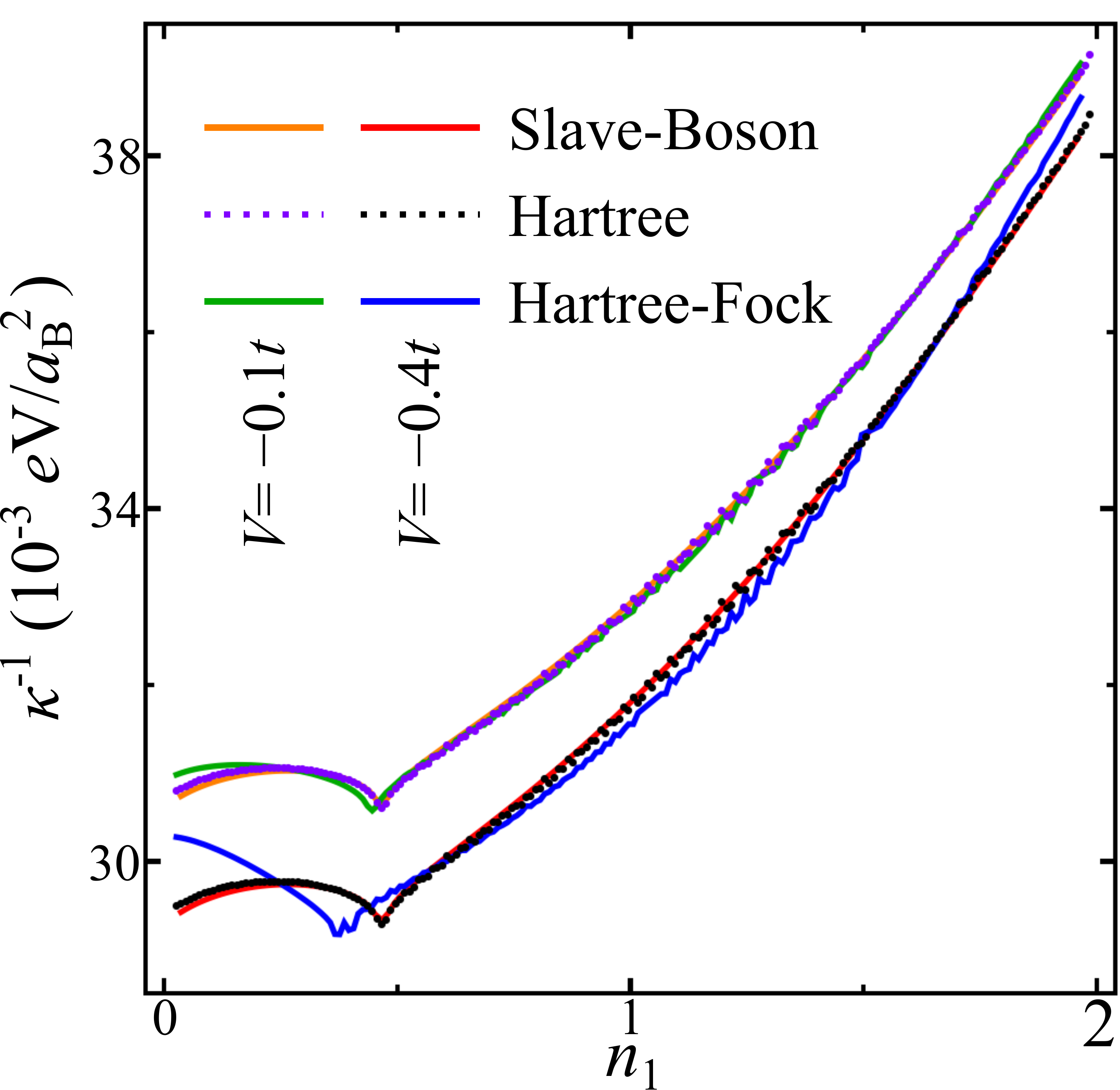}
\caption
{(Color online) The difference of the inverse compressibilities for the system described in Fig.~\ref{U1metalSBHFCap} is largest for low densities and around the vHs filling.}
\label{U1metalSBHFComp}
\end{figure}

In experimental setups the surface electrode is  typically distinct and comprises a weakly correlated electron system. Here, we consider a heterostructure with an uncorrelated electron system at the surface electrode (labeled  ``2'') whereas the interface electrode (labeled ``1'') comprises a strongly correlated electron
system. 

For plate 2 (surface), a two-dimensional metal with free electrons of effective mass ratio $m_2/m_e$ yields
\begin{align}
F_2(n_{\rm tot}-n_1)&=\frac{\pi a_{\rm B}e^2}{m_2/m_e}(n_{\rm tot}-n_1)^2
\end{align}
for the free energy, where $a_{\rm B}$ is the bare Bohr radius. We take ${m_2/m_e}$ to be 5 throughout this section. A (more realistic) effective Bohr radius can be integrated into a modified effective mass $m_2$. For plate 1 (interface) we take strongly correlated electrons with $U=9\,t$ and $t'=-0.15\,t$.

The second derivative with respect to $n_1$ does not depend on the density,
\begin{align}
F_2''&=\frac{2\pi a_{\rm B}e^2}{m_2/m_e},
\end{align}
so that the only density-dependent quantity in the compressibility and capacitance enters via $F_1''$ (c.f. Eq.~(\ref{eq:secondderivativelayout}) and Eq.~(\ref{eq:differentialcapacitance2})). This is reflected in Figs.~\ref{U9metalComp} and \ref{U9metalCap}, where we analyzed the system described in Fig.~\ref{U9} and replaced the surface by a two-dimensional metal with free electrons. The total electron density is fixed to $n_{\rm tot}=2.0$. The more complex structure of a double peak and double dip around $n_1=n_{\rm tot}/2$ present in Fig.~\ref{U9}~(e) and (h), respectively,  is not observed for the asymmetric system of this section; the structure 
of the inverse compressibility in Fig.~\ref{U9metalComp} is more reminiscent to that of Fig.~\ref{U9}~(b).
The capacitance in Fig.~\ref{U9metalCap} is normalized to $C_{U=0,V=0}$ and shows clearly the ``correlation dip'' when the density of the interface electrons is close to half filling.

The capacitance for an extended Hubbard system with $U=t$ and $t'=-0.45\,t$ coupled to a two-dimensional free electron gas is shown in Fig.~\ref{U1metalSBHFCap}. As before, since the contribution of $F_2''$ to the capacitance is constant, the density dependence is dominated by $F_1''$ (c.f. Fig.~\ref{U1}\,(c)). The normalization to the geometric capacitance reveals a slight increase of $C$ for $V=-0.4\,t$ above $C_{\rm geo}=4\pi d/\varepsilon$ when the system is at one of the van Hove singularities for fillings $n_1=0$ or $n_1\approx 0.47$. Elsewhere the compressibility of the single layer $\kappa_0$ is positive and hence the capacitance is reduced below the geometric value. The compressibility of the total system (Fig.~\ref{U1metalSBHFComp}) is approximately that of Fig.~\ref{U1}\,(c), up to a constant.

Fig.~\ref{U1metalSBHFCap} and Fig.~\ref{U1metalSBHFComp} include also a comparison between the slave-boson, Hartree and Hartree-Fock approximations. The slave-boson technique agrees excellently with the Hartree calculations. The difference between Hartree and Hartree-Fock is due to the second derivative of the Fock term. This contribution is most notable for empty and full bands, and at the vHs at intermediate filling. 


\section{Compressibility and Capacitance}\label{appendix_C}

For the derivation of a general relation between total compressibility and differential capacitance we assume two conditions to be fulfilled: Firstly, the free energy of the total system can be written in the form of Eq.~(\ref{eq:freeenergy}),
\begin{align}\label{app:freeenergy}
F(n_{\rm t},n_1)&=F_1(n_1)+F_2(n_{\rm t}-n_1)+F_{\rm es}(n_{\rm t},n_1)\notag\\
&\qquad\qquad-eV_{\rm ext}n_1A,
\end{align}
where we introduced the abbreviation $n_{\rm t}\equiv n_{\rm tot}$ for this section. Secondly, the internal variable $n_1$ minimizes the free energy and is not a boundary value, i.e., $n_1\neq 0$ and $n_1\neq n_{\rm t}$. This stays valid for a differential change of the total charge $n_{\rm t}$:
\begin{align}\label{app:minimum1}
\frac{\D}{\D n_{\rm t}}\frac{\partial F}{\partial n_1}&=0 \\
\Longleftrightarrow\frac{\partial^2F}{\partial n_1\partial n_{\rm t}} &= -\frac{\partial^2F}{\partial n_1^2}\frac{\partial n_1}{\partial n_{\rm t}}\label{app:minimum2}
\end{align}
The inverse compressibility of the system is proportional to the second derivative with respect to the total charge density:
\begin{align}
\kappa^{-1}A/n_{\rm t}^2&=\frac{\D^2F}{\D n_{\rm t}^2}\\
&=\frac{\partial^2F}{\partial n_{\rm t}^2}+2\frac{\partial^2F}{\partial n_1\partial n_{\rm t}}\frac{\partial n_1}{\partial n_{\rm t}}+\frac{\partial^2F}{\partial n_1^2} \left(\frac{\partial n_1}{\partial n_{\rm t}}\right)^2\notag\\
&\overset{(\ref{app:minimum2})}{=}\frac{\partial^2F}{\partial n_{\rm t}^2}-\frac{\partial^2F}{\partial n_1^2}\left(\frac{\partial n_1}{\partial n_{\rm t}}\right)^2
\label{app:comp}
\end{align}
We make use of the form of the free energy,
\begin{align}
\frac{\partial^2F}{\partial n_1\partial n_{\rm t}}&=-F_2''+\frac{\partial^2F_{\rm es}}{\partial n_1\partial n_{\rm t}}\overset{(\ref{app:minimum2})}{=}-\frac{\partial^2F}{\partial n_1^2}\frac{\partial n_1}{\partial n_{\rm t}}\label{app:mixedpartial}\\
\frac{\partial^2F}{\partial n_{\rm t}^2}&=F_2''+\frac{\partial^2F_{\rm es}}{\partial n_{\rm t}^2}\notag\\
&\overset{(\ref{app:mixedpartial})}{=}\frac{\partial^2F}{\partial n_1^2}\frac{\partial n_1}{\partial n_{\rm t}} +\frac{\partial^2F_{\rm es}}{\partial n_1\partial n_{\rm t}} +\frac{\partial^2F_{\rm es}}{\partial n_{\rm t}^2}\label{app:doublepartial}.
\end{align}
The last line is inserted in the relation for the compressibility (\ref{app:comp}):
\begin{align}\label{app:compress}
\frac{A}{n_{\rm t}^2}\kappa^{-1}&=\frac{\partial^2F}{\partial n_1^2}\frac{\partial n_1}{\partial n_{\rm t}}\left(1-\frac{\partial n_1}{\partial n_{\rm t}}\right)+\left(\frac{\partial^2}{\partial n_{\rm t}^2}+\frac{\partial^2}{\partial n_1\partial n_{\rm t}}\right)F_{\rm es}\notag\\
&= \frac{e^2A^2}{C_{\rm diff}}\frac{\partial n_1}{\partial n_{\rm t}}\left(1-\frac{\partial n_1}{\partial n_{\rm t}}\right)+\mathcal{D}F_{\rm es},
\end{align}
where we introduced the differential operator  $\mathcal{D}\equiv \left(\partial_{n_{\rm t}}^2+\partial_{n_1}\partial_{n_{\rm t}}\right)$. This relation corresponds to
Eq.~(\ref{compressibilitychargetransfer}) in Sec.~\ref{sec:basics}.

A generalized distribution of the positive background charge is depicted in Fig.~\ref{GeneralizedSetup}. We assume that the distance between the electrodes is given by $d$. The fraction $m_1$ of the total positive charge resides at distance $\alpha d$ from the lower electrode, while the rest of the positive charge is 
at a distance $\beta d$ from the lower electrode. For $\alpha=\beta=0.5$ we recover the polar heterostructure of Fig.~\ref{Setup} (with $d^l_a=d^l_b$ and 
$\varepsilon_a=\varepsilon_a$), and for $\alpha=0$, $\beta=1$ the standard configuration where the layers of positive charge are identical with the layers of negative charge. The electrostatic energy of the general layout is
\begin{align}\label{app:generallayout}
F_{\rm es}^g&=2D\big[\alpha n_1^2+(\beta-\alpha)\left(m_1n_{\rm t}-n_1\right)^2\notag\\
&\qquad +(1-\beta)\left(n_{\rm t}-n_1\right)^2\big],
\end{align}
with $D=\pi e^2 A d/\varepsilon$. This generalized model allows to analyze the effect of a variety of electrostatic configurations on the compressibility in one framework. The respective contribution to the differential capacitance,
\begin{align}
\frac{\partial^2F_{\rm es}^g}{\partial n_1^2}&=\frac{e^2A}{C_{\rm geo}}=4 D,
\end{align}
is independent of the layout of the positive charge and determined by the distance between the electrodes. The differential operator,
\begin{align}\label{app:diffopgeneral}
\mathcal{D}F_{\rm es}^g&=4D\left(\beta-\alpha\right)m_1\left(m_1-1\right),
\end{align}
is zero if all positive charge resides in one plane ($\alpha=\beta$) and otherwise negative. These two results can be combined with Eq.~(\ref{app:compress}),
\begin{align}\label{app:stablesystem}
\frac{\kappa^{-1}A}{n_{\rm t}^2}&=\frac{e^2A^2}{C_{\rm diff}}\frac{\partial n_1}{\partial n_{\rm t}}\left(1-\frac{\partial n_1}{\partial n_{\rm t}}\right)\notag\\
&\qquad-\frac{e^2A^2}{C_{\rm geo}}\left(\beta-\alpha\right)m_1\left(1-m_1\right).
\end{align}
Eq.\,(\ref{app:stablesystem}) yields the condition for a system to be stable ($\kappa>0$):
\begin{align} 
\frac{C_{\rm diff}}{C_{\rm geo}}\left(\beta-\alpha\right)m_1\left(1-m_1\right)&<\frac{\partial n_1}{\partial n_{\rm t}}\left(1-\frac{\partial n_1}{\partial n_{\rm t}}\right)\label{app:limit}.
\end{align}
This relation limits the enhancement of the differential capacitance over its geometric value, since the right hand side of the last inequality is $\leq 1/4$. 

We now select three special cases for the distribution of positive background charge:
\begin{itemize}
\item $\beta^a=\alpha^a$\\
The asymmetric case $a$, where the whole positive charge is concentrated in one layer,
\item $\beta^s=1-\alpha^s$, $m^s_1=1/2$\\
the symmetric distribution $s$ of the positive charge and
\item $\alpha^t=0$, $\beta^t=1$, $0\leq m_1^t\leq1$\\\
the standard configuration $t$ with the planes of positive charge coinciding with the electrodes.
\end{itemize}
For $\alpha^a=\alpha^s=1/2$ the first two cases are equal and recover the \textit{symmetric polar heterostructure} (SPH) introduced in section \ref{sec:layout}. The \textit{symmetric standard capacitor} (SSC) is obtained for $\alpha^s=0$ and $m_1^t=1/2$ for the last two cases and the \textit{asymmetric standard capacitor} (ASC) for $\alpha^a=1$ and $m_1^t=0$.

For the asymmetric layout $a$ the differential operator (\ref{app:diffopgeneral}) vanishes so that there is no limitation to the capacitance enhancement. In the standard configuration, on the other hand, a necessary condition for its stability is given by
\begin{align}
\frac{C_{\rm diff}^t}{C_{\rm geo}}\leq\frac{1}{4m_1^t\left(1-m_1^t\right)}.
\end{align}
This implies that for $m^t_1=1/2$, which is the symmetrical standard configuration, no enhancement above the geometrical capacitance is possible. The more asymmetric the positive charge is distributed, the larger the allowed capacitance enhancement is.

The differential capacitance of the symmetric configuration is limited by
\begin{align}
\frac{C_{\rm diff}^s}{C_{\rm geo}}\leq\frac{1}{1-2\alpha^s}.
\end{align}
In this layout the condition used to determine $n_1$,
\begin{align}
\partial_{n_1} F^s&= \partial_{n_1}F_1+\partial_{n_1}F_2+4D\left(n_1-\frac{n_{\rm t}}{2}\right)-eV_{\rm ext}A\overset{!}{=}0,\notag
\end{align}
is independent of $\alpha^s$. Hence the solution $n_1^s$ and the charge transfer function $\partial_{n_{\rm t}} n_1^s$ are equal for all $\alpha^s$. Note that, due to
\begin{align}
\mathcal{D}F_{\rm es}^s&=-D\left(1-2\alpha^s\right)
\end{align}
and Eq.~(\ref{app:compress}), the inverse compressibilities for different symmetric layouts differ by a constant proportional to $1-2\alpha^s$. The nearer the positive charge is to the electrodes, the more compressible the total system becomes.

Finally, we derive Eq.~(\ref{compressibilitysimplified}) which specifies the compressibility for the SPH, SSC and ASC layouts.
Firstly, we can replace $\partial n_1/\partial n_{\rm t}$ in Eq.~(\ref{app:comp}) by Eq.~(\ref{app:minimum2}),
\begin{align}
\frac{\D^2 F}{\D n_{\rm t}^2}&=\frac{\partial^2F}{\partial n_{\rm t}^2}-\left(\frac{\partial^2F}{\partial n_1\partial n_{\rm t}}\right)^2\bigg/\frac{\partial^2F}{\partial n_1^2}\notag\\
&=\left(\frac{\partial^2F}{\partial n_1^2}\right)^{-1}\left[\frac{\partial^2F}{\partial n_1^2}\frac{\partial^2F}{\partial n_{\rm t}^2}-\frac{\partial^2F}{\partial n_1\partial n_{\rm t}}\right]
\end{align}
and then make use of the special form of the free energy, Eq.~(\ref{eq:freeenergy}):
\begin{align}\label{app:specialconfigs}
\frac{\D^2 F}{\D n_{\rm t}^2}&=\frac{1}{F_1''+F_2''+4D}\bigg[(F_1''+F_2''+4D)\left(F_2''+ \frac{\partial^2F_{\rm es}}{\partial n_{\rm t}^2}\right)\notag\\
&\qquad-\left(-F_2''+\frac{\partial^2F_{\rm es}}{\partial n_1\partial n_{\rm t}}\right)^2\bigg],
\end{align}
where $F_{1,2}''=\partial^2F_{1,2}/\partial n_1^2$. The electrostatic energies for the different layouts are
\begin{align}
F_{\rm es}^{\rm SPH}&=D[n_1^2+(n_{\rm t}-n_1)^2]\notag\\
F_{\rm es}^{\rm SSC}&=2D\left(\frac{n_{\rm t}}{2}-n_1\right)^2\\
F_{\rm es}^{\rm ASC}&=2Dn_1^2\notag
\end{align}
and the corresponding derivatives yield
\begin{align}
\frac{\partial^2F_{\rm es}^{\rm SPH}}{\partial n_{\rm t}^2}&=2D, &\frac{\partial^2F_{\rm es}^{\rm SPH}}{\partial n_1\partial n_{\rm t}}=-2D\notag\\
\frac{\partial^2F_{\rm es}^{\rm SSC}}{\partial n_{\rm t}^2}&=D, &\frac{\partial^2F_{\rm es}^{\rm SSC}}{\partial n_1\partial n_{\rm t}}=-2D\\
\frac{\partial^2F_{\rm es}^{\rm ASC}}{\partial n_{\rm t}^2}&=0, &\frac{\partial^2F_{\rm es}^{\rm ASC}}{\partial n_1\partial n_{\rm t}}=0.\notag
\end{align}
Inserting these expressions into Eq.~(\ref{app:specialconfigs}) yields Eq.~(\ref{compressibilitysimplified}).

Systems with $F_1''>0$ and $F_2''>0$ are always stable, irrespective of the electrostatic layout: 

Substitution of the general electrostatic energy $F_{\rm es}^g$ of Eq.~(\ref{app:generallayout}) in Eq.~(\ref{app:specialconfigs}), with the partial derivatives
\begin{align}
\frac{\partial^2F_{\rm es}^g}{\partial n_1\partial n_{\rm t}}=-4D\left[(\beta-\alpha)m_1+(1-\beta)\right]\equiv -4D\Delta\notag\\
\frac{\partial^2F_{\rm es}^g}{\partial n_{\rm t}^2}=4D\left[(\beta-\alpha)m_1^2+(1-\beta)\right]\equiv 4D\Gamma\notag
\end{align}
yields
\begin{align}
\frac{\D^2 F}{\D n_{\rm t}^2}&=\frac{1}{F_1''+F_2''+4D}\bigg[(F_1''+F_2''+4D)\left(F_2''+ 4D\Gamma\right)\notag\\
&\qquad-\left(-F_2''-4D\Delta\right)^2\bigg].
\end{align}
Hence the compressibility of the total system has the same sign as the expression
\begin{align}
(F_1''&+F_2''+4D)\left(F_2''+ 4D\Gamma\right)-\left(F_2''+4D\Delta\right)^2\notag\\
&=4DF_2''\left(\Gamma+1-2\Delta\right)+(4D)^2\left(\Gamma-\Delta^2\right)\notag\\
&\quad+F_1''\left(F_2''+ 4D\Gamma\right)
\end{align}
The last summand is positive, since $\Gamma>0$. For the other two terms we find:
\begin{align}
\Gamma+1-2\Delta&=(\beta-\alpha)(m_1-1)^2+\alpha>0\notag\\
\Gamma-\Delta^2&=\alpha\left[(\beta-\alpha)m_1+(1-\beta)\right]\notag\\
&\qquad+(\beta-\alpha)(1-\beta)(1-m_1)^2>0\notag
\end{align}
so that the total expression is always positive.
Hence, we conclude that  the total compressibility is always positive provided that the
compressibilities of the subsystems---the two electrodes---are positive.

\end{document}